\documentclass[%
 aip,
 amsmath,amssymb,
reprint,
]{revtex4-1}

\usepackage[colorlinks=true, linkcolor=blue, citecolor=blue]{hyperref}
\usepackage{graphicx}
\usepackage{color}

\draft % marks overfull lines with a black rule on the right

\begin{document}

%\preprint{}

\title{Model of pulsar pair cascades in non uniform electric fields: growth rate, density profile and screening time} %Title of paper



\author{F. Cruz}
\email[]{fabio.cruz@tecnico.ulisboa.pt}
\author{T. Grismayer}
\affiliation{
 GoLP/Instituto de Plasmas e Fus\~{a}o Nuclear, \\ Instituto Superior T\'{e}cnico, Universidade de Lisboa, 1049-001 Lisboa, Portugal
}
\author{S. Iteanu}
\affiliation{
Département de Physique Univ. Lyon, ENS de Lyon, Univ. Claude Bernard, F-69342 Lyon, France
}
\author{P. Tortone}
\affiliation{
 Dipartimento di Energia, Politecnico di Torino, 10129 Torino, Italy
}
\author{L. O. Silva}
\email[]{luis.silva@tecnico.ulisboa.pt}
\affiliation{
 GoLP/Instituto de Plasmas e Fus\~{a}o Nuclear, \\ Instituto Superior T\'{e}cnico, Universidade de Lisboa, 1049-001 Lisboa, Portugal
}

% todo: check affiliations

\date{\today}

\begin{abstract}
Time-dependent cascades of electron-positron pairs are thought to be the main source of plasma in pulsar magnetospheres and a primary ingredient to explain the nature of pulsar radio emission, a longstanding open problem in high-energy astrophysics. During these cascades --- positive feedback loops of gamma-ray photon emission, via curvature radiation by TeV electrons and positrons, and pair production ---, the plasma self-consistently develops inductive waves that couple to electromagnetic modes capable of escaping the pulsar dense plasma. In this work, we present an analytical description of pair cascades relevant in pulsars, including their onset, exponential growth and saturation stages. We study this problem in the case of a background linear electric field, relevant in pulsar polar caps, and using an heuristic model of the pair production process. The analytical results are confirmed with particle-in-cell simulations performed with OSIRIS including heuristic pair production.
\end{abstract}


\maketitle %\maketitle must follow title, authors, abstract and \pacs

\section{Introduction}
\label{sec:introduction}

% extreme magnetic fields > magnetosphere > deviations from force-free
Pulsars are rapidly rotating neutron stars permeated by magnetic fields that can exceed $10^{12}$~G. Their exotic magnetospheres are filled with highly relativistic pair-dominated magnetized plasma flows that transport the rotational energy of the neutron star. These flows are such that the pulsar magnetosphere is in the force-free regime almost everywhere~\citep{michel_1982, petri_2016, cerutti_2017}. This regime can be expressed by the condition $\mathbf{E} \cdot \mathbf{B} \simeq 0$, where $\mathbf{E}$ and $\mathbf{B}$ are the ambient electric and magnetic fields. Regions where this condition cannot be met are ideally suited for nonlinear phenomena, such as: i) dissipation via e.g. magnetic reconnection~\citep{coroniti_1990}, leading to high-energy gamma-ray emission~\citep{lyubarskii_1996}, and ii) pair cascades~\citep{sturrock_1971, ruderman_sutherland_1975}, a feedback loop that supplies plasma to the magnetosphere.

% cascades: physical processes, locations
In pair cascades, electrons and positrons are accelerated along the curved magnetic field of pulsars, emitting gamma-ray photons through curvature radiation that can be converted into new electron-positron pairs via Quantum Electrodynamics (QED) processes~\citep{erber_1966, ritus_1985}. If emitting and newly-born leptons are continuously accelerated, this mechanism can act as an efficient source of pair plasma that self-regulates~\citep{sturrock_1971, ruderman_sutherland_1975}: it stops when the fresh plasma is dense enough to screen the background electric field, and restarts if/when the plasma advects into other regions of the magnetosphere.

% gaps, works on ab initio cascades
Vacuum gaps, \textit{i.e.}, regions of $\mathbf{E} \cdot \mathbf{B} \neq 0$ where cascades develop, can be located in polar caps\citep{sturrock_1971, ruderman_sutherland_1975}, near the stellar surface or in the outer magnetosphere~\citep{arons_1983, cheng_1986}. Cascades operating at low altitudes have been conjectured to be intimately connected to coherent emission mechanisms from pulsars~\citep{sturrock_1971, ruderman_sutherland_1975}. Recent advances in numerical algorithms and computational resources allowed a great progress in the understanding of pair cascades and their interplay with collective plasma phenomena. \citet{timokhin_2010} performed the first 1D simulations including the relevant QED effects from first principles in plasma kinetic simulations, showing that cascades periodically launch large-density plasma blobs into the magnetosphere, generating coherent superluminal electrostatic waves~\citep{levinson_2005}. Follow-up works~\cite{timokhin_2013, timokhin_2015} determined the spectra and multiplicity of the pair plasma created in cascades for a variety of initial conditions. Recently, 2D simulations have shown that the modes identified in 1D simulations can be electromagnetic if the pair production fronts are not perfectly aligned with the local magnetic field~\cite{philippov_2020, cruz_2021b}. These results demonstrated that the properties of pulsar radio emission can be intimately connected with the QED and plasma kinetic processes underlying pair cascades.

% motivation: lack of theoretical understanding of heuristics, works on heuristic models
Exploring the intricate interplay between these processes with \textit{ab initio} simulations is computationally very expensive. The large multiplicity of cascades and the consequent disparity between the size of the vacuum gaps ($\sim 100$~m) and the shortest plasma kinetic scales ($\sim 1$~cm) make 2D and 3D simulations very challenging, even when modelling only local simulation domains. Heuristic models of the QED processes emerged as a natural solution to this problem. These models replace the strong nonlinearities of the QED cross sections with simpler criteria, with the goal of lowering the computational costs of simulations of pair cascades.

A commonly adopted heuristic model consists in injecting electron-positron pairs at a rate that depends on the local electric and/or magnetic fields. This strategy has been used as a plasma source in global simulations of pulsar magnetospheres~\citep{philippov_2014, belyaev_2015, cerutti_2015, kalapotharakos_2018, brambilla_2018} to study the distribution of plasma currents in the vicinity of the neutron star, as well as particle acceleration in the current sheets that develop beyond the light cylinder, leading to gamma-ray emission consistent with observations. A second class of heuristic models replaces the criteria for QED processes to occur with thresholds based on typical energy or length scales associated with these processes. Such models are able to trace the development of pair cascades, and thus capture more accurately the interaction between QED and plasma kinetic processes. These models have been used in local simulations of pair cascades~\citep{levinson_2005, philippov_2020, cruz_2021}, but also in global pulsar magnetosphere simulations~\citep{chen_2014, philippov_2015a, philippov_2015b, chen_2020, guepin_2020}.

Despite their wide use and known mapping to first-principles descriptions of the QED processes in some cases~\citep{cruz_2021}, heuristic models remain poorly explored from a theoretical standpoint. These simple models are an important tool to understand the development of pair cascades and to rigorously derive estimates for e.g. their growth rate, the plasma distribution functions that they produce and the types and properties of plasma waves excited. Such estimates are important to comprehend coherent emission processes triggered during the cascade process, but also other processes that depend on cascade products, namely the properties of the plasma distribution function~\citep{benacek_2021a, benacek_2021b}.

% paper outline
In this work, we use an energy threshold-based heuristic model to analytically and numerically study the development of pair cascades. In Sec.~\ref{sec:uniform}, we describe this model in detail and assume, for simplicity, that the cascade develops in a uniform background electric field to derive its properties. In Sec.~\ref{sec:linear}, we investigate how are the properties of the cascade modified with a more realistic linear electric field profile, in particular how is its growth rate modified and how it changes in time and space. In Secs.~\ref{sec:uniform} and \ref{sec:linear}, we compare our analytical results with particle-in-cell (PIC) simulations. Our conclusions are finally presented in Sec.~\ref{sec:conclusions}.

\section{Cascade in a uniform electric field}
\label{sec:uniform}

Polar cap pair discharges are regulated by two QED processes: i) emission of gamma-ray photons by ultra-relativistic electrons and positrons via nonlinear Compton scattering~\citep{erber_1966}, the QED equivalent of curvature radiation~\citep{kelner_2015, delgaudio_2020}, and ii) conversion of gamma-rays into electron-positron pairs via multiphoton Breit-Wheeler pair production~\citep{ritus_1985}. The differential probability rates for these processes are well-known~\citep{erber_1966, ritus_1985} functions of the parent particles' energy, momentum components and local electromagnetic field components.

However, for typical pulsar parameters (surface magnetic field of $B \sim 10^{12}$~G and a rotation period of $T \sim 1$~s), we can estimate the characteristic energy and length/time scales associated with these processes in polar caps. For instance, the energy at which electrons and positrons produce the most energetic photons is $\varepsilon_\pm \sim 10^7~m_e c^2$, where $m_e$ is the electron mass and $c$ is the speed of light. The length scale associated with this process is the distance over which leptons are accelerated from rest to these energies, $\ell_\pm \sim 100$~m. The photons emitted by the ultra-relativistic leptons follow an energy spectrum that peaks at the critical energy $\varepsilon_\gamma \sim 10^2 - 10^3~m_e c^2$, and decay into pairs after a typical propagation distance $\ell_\gamma \sim 1-10$~m.

In this work, we assume that these processes can be described by the following heuristic model of pair production: any electron or positron accelerated beyond a threshold energy $\varepsilon_\mathrm{thr}$ emits a new electron-positron pair with a fraction $f$ of its original energy, $\varepsilon_\mathrm{pair} \equiv f \varepsilon_\mathrm{thr}$, equally split between the pair particles. With this model, we assume that the intermediate photon decays instantaneously, \textit{i.e.}, $\ell_\gamma \ll \ell_\pm$, and that the QED processes occur at well-defined energies. These conditions are valid for the canonical parameters presented above. However, they become invalid e.g. for pulsars with lower periods or weaker surface magnetic fields, for which the ratio $\ell_\gamma / \ell_a$ can be significantly different and a one-dimensional description of the cascade is likely insufficient~\citep{cruz_2021b}.

For convenience, we parametrize the model presented in this work using the fraction $f$ and the threshold Lorentz factor $\gamma_\mathrm{thr} = \varepsilon_\mathrm{thr} / m_e c^2$. In real pulsars, $\gamma_\mathrm{thr} \sim \varepsilon_\pm / m_e c^2 \sim 10^7$ and $f \sim 10^{-6} - 10^{-3} \ll 1$, although exact estimates may depend on the magnitude and geometry of the surface magnetic field of the pulsar~\cite{cruz_2021}.

In general, cascades develop in electric fields with spatial dependencies. For example, in polar cap vacuum gaps, the electric field varies both in altitude and latitude, leading to non trivially shaped pair production fronts~\cite{arons_1979, philippov_2020} and emission power profiles~\cite{cruz_2021b}. For simplicity, we assume in this work that the cascade can be described as a 1D problem along the magnetic field lines. This is a good approximation due to the strong magnetic fields that permeate pulsar magnetospheres, which forbids charged particles to cross magnetic field lines. In this Section, we also assume that the cascade develops on a spatially uniform electric field with magnitude $E_0$. We investigated this scenario in detail in \citet{cruz_2021}. In this Section, we review the key results from that work in order to introduce notation and establish the theoretical framework required in the more complex scenario studied in Sec.~\ref{sec:linear}.

\subsection{Population model}
\label{sec:population-model}

\begin{figure}
\centering
\includegraphics[width=2.9in]{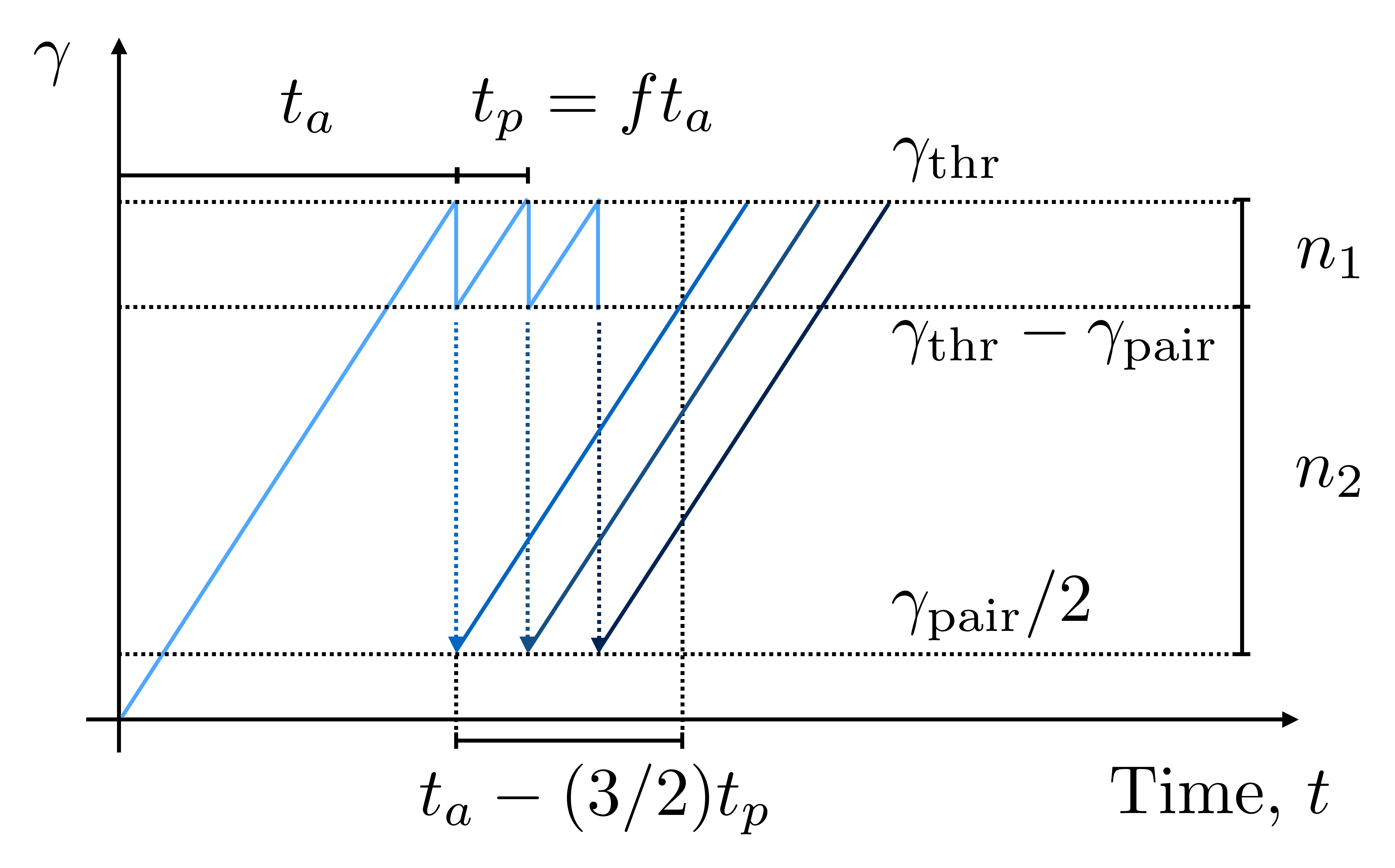}
\caption{\label{fig:scheme-cascade} Schematic representation of heuristic pair production model and associated populations and time scales. Solid blue lines represent the time evolution of the Lorentz factor of electrons (or positrons). When electrons reach a Lorentz factor $\gamma_\mathrm{thr}$, they emit a pair with energy $\gamma_\mathrm{pair} m_e c^2$. Emission times are indicated with blue dashed arrows, and new particles created at those times are shown with darker colors. The energy bands, corresponding to populations 1 and 2 described in the text, are indicated on the right hand side. Characteristic times $t_a$ and $t_p$ are also schematically indicated.}
\end{figure}

It is straightforward to determine the two relevant time scales of this system with electrons and positrons subject to a uniform electric field $E_0$ pair producing according to the heuristic model presented above: i) $t_a \equiv \gamma_\textrm{thr} m_e c / e E_0$ is the time required for particles to be accelerated from rest to $\gamma_\mathrm{thr}$, and ii) $t_p \equiv f t_a$ is the time required for particles that just emitted a pair to be reaccelerated to $\gamma_\mathrm{thr}$. These time scales are illustrated in the schematic representation of the cascade process in Fig.~\ref{fig:scheme-cascade}.

Any particle in this simple system emits new pairs every $t_p$ after being accelerated to $\gamma_\mathrm{thr}$ for the first time. This naturally divides the system in two populations: $n_1$ particles with $\gamma \in [\gamma_\mathrm{thr}-\gamma_\mathrm{pair},  \gamma_\mathrm{thr}]$, and $n_2$ particles with $\gamma < \gamma_\mathrm{thr}-\gamma_\mathrm{pair}$ (see Fig.~\ref{fig:scheme-cascade}). Particles in population 2 are created with a Lorentz factor $f \gamma_\mathrm{thr} / 2$ and convert into population 1 over a time $t \simeq t_a - 3 t_p / 2 = t_a (1 - 3 f /2)$, \textit{i.e.}, we can write
\begin{equation}
n_1 (t + t_a (1 - 3 f / 2)) \simeq n_1(t) + n_2(t) \ .
\label{eq:cascade_delay}
\end{equation}
Symmetrically, particles in population 2 decrease at a rate $n_2 / (t_a (1 - 3 f / 2))$ due to this conversion, and increase at a rate $2 n_1 / t_a$ due to pair production from both species, and we can write
\begin{equation}
\frac{\mathrm{d} n_2(t)}{\mathrm{d} t} \simeq \frac{2 n_1(t)}{f t_a} - \frac{n_2(t)}{(1-3f/2)t_a} \ .
\label{eq:cascade_deriv}
\end{equation}

For $f \ll 1$, Eqs.~\eqref{eq:cascade_delay} and \eqref{eq:cascade_deriv} have an exponentially growing solution $n_{1, 2} (t) \propto \exp(\Gamma t)$, with
\begin{equation}
\Gamma t_a \simeq W(2 / f) \simeq \ln (2 / f) \ ,
\label{eq:gamma_uniform}
\end{equation}
where $W(x)$ is the Lambert function~\citep{corless_1996}, which behaves as $\ln(x)$ for large $x$. We note that this solution for $\Gamma$ scales very weakly with $f$ for all relevant $f \ll 1$.

To verify our theoretical analysis, we have performed 1D PIC simulations with OSIRIS~\citep{fonseca_2002, fonseca_2008} of a uniform electron-positron plasma initially at rest and immersed in a uniform electric field $E_0$. The value of $E_0$ is chosen such that particles attain relativistic energies over short distances, \textit{i.e.}, $e E_0 (c / \omega_p) / m_e c^2 \simeq 3000 \gg 1$, where $\omega_p = (4 \pi e^2 n_0 / m_e)$ is the plasma frequency associated with the initial plasma density $n_0$ and $e$ is the electron charge. The simulation domain has length $L / (c / \omega_p) \simeq 30$ and periodic boundary conditions. The grid resolution is $\Delta x / (c / \omega_p) = 0.015$, and the time step is chosen to resolve $t_p$, the shortest time scale of the problem. In the PIC simulations, pair production is included according to the energy threshold-based heuristic model described above.

\begin{figure}
\includegraphics[width=3.4in]{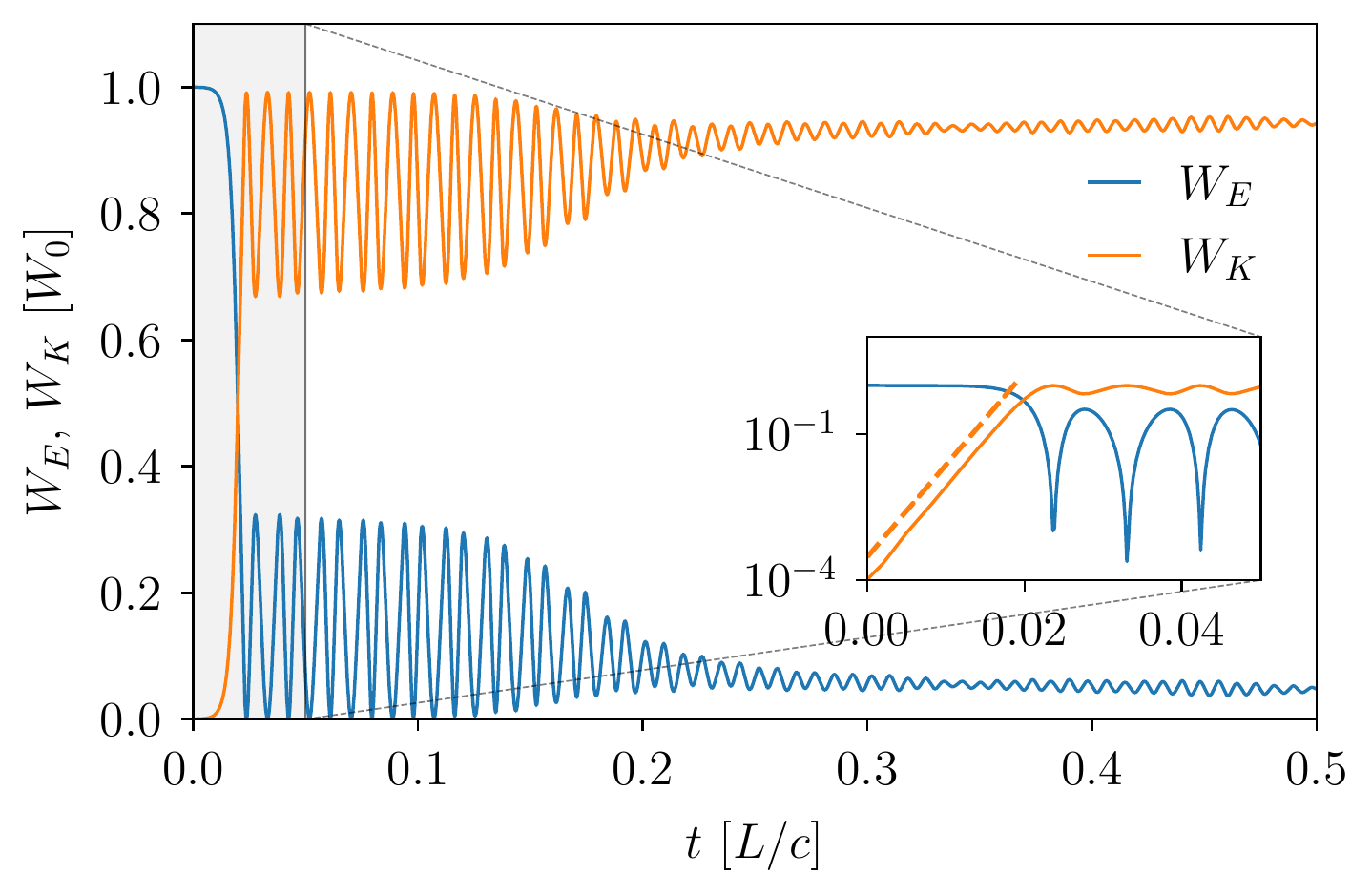}
\caption{\label{fig:const-ene} Time evolution of the energy stored in the electric field $W_E$ and of the total kinetic energy $W_K$ in a simulation of a pair cascade with $\gamma_\mathrm{thr} = 500$ and $f = 0.1$ developed in a uniform electric field $E_0$. The energies $W_E$ and $W_K$ are normalized to the total energy available at time $t = 0$, $W_0 = \int_0^L (E_0^2 / 8 \pi) \ \mathrm{d} x$.}
\end{figure}

In Fig.~\ref{fig:const-ene}, we represent the evolution in time of the energy in the electric field $W_E = \int_0^L (E^2 / 8 \pi) \ \mathrm{d} x$ and of the total kinetic energy $W_K = \sum_i (\gamma_i - 1) m_e c^2$ (where the sum is done over all particles) for a simulation with $\gamma_\mathrm{thr} = 500$ and $f = 0.1$. In the inset, we highlight the exponential growth of the kinetic energy in the early times of the simulation and show that it is well described by a growth rate $\Gamma$ given by Eq.~\eqref{eq:gamma_uniform}. At these early times, the total kinetic energy in the system grows due to a growing number of particles, whereas the average kinetic energy is approximately constant and is controlled by the distribution function developed during the cascade~\citep{cruz_2021}.

\subsection{Inductive oscillations}
\label{sec:inductive}

As new pairs are created during the cascade, electrons and positrons are accelerated in opposite directions, driving a current. Due to the spatial homogeneity of $E$, the density of electrons and positrons $n_\pm$ is constant in space (and equal in magnitude) during the cascade, and so is the current $j$ that they drive. As the current grows in time, the electric field is screened, as prescribed by Amp\`{e}re's law
\begin{equation}
\frac{\partial E}{\partial t} = -4 \pi j \simeq 8\pi e c n_\pm \ .
\label{eq:ampere_screen}
\end{equation}

\begin{figure}
\includegraphics[width=3.4in]{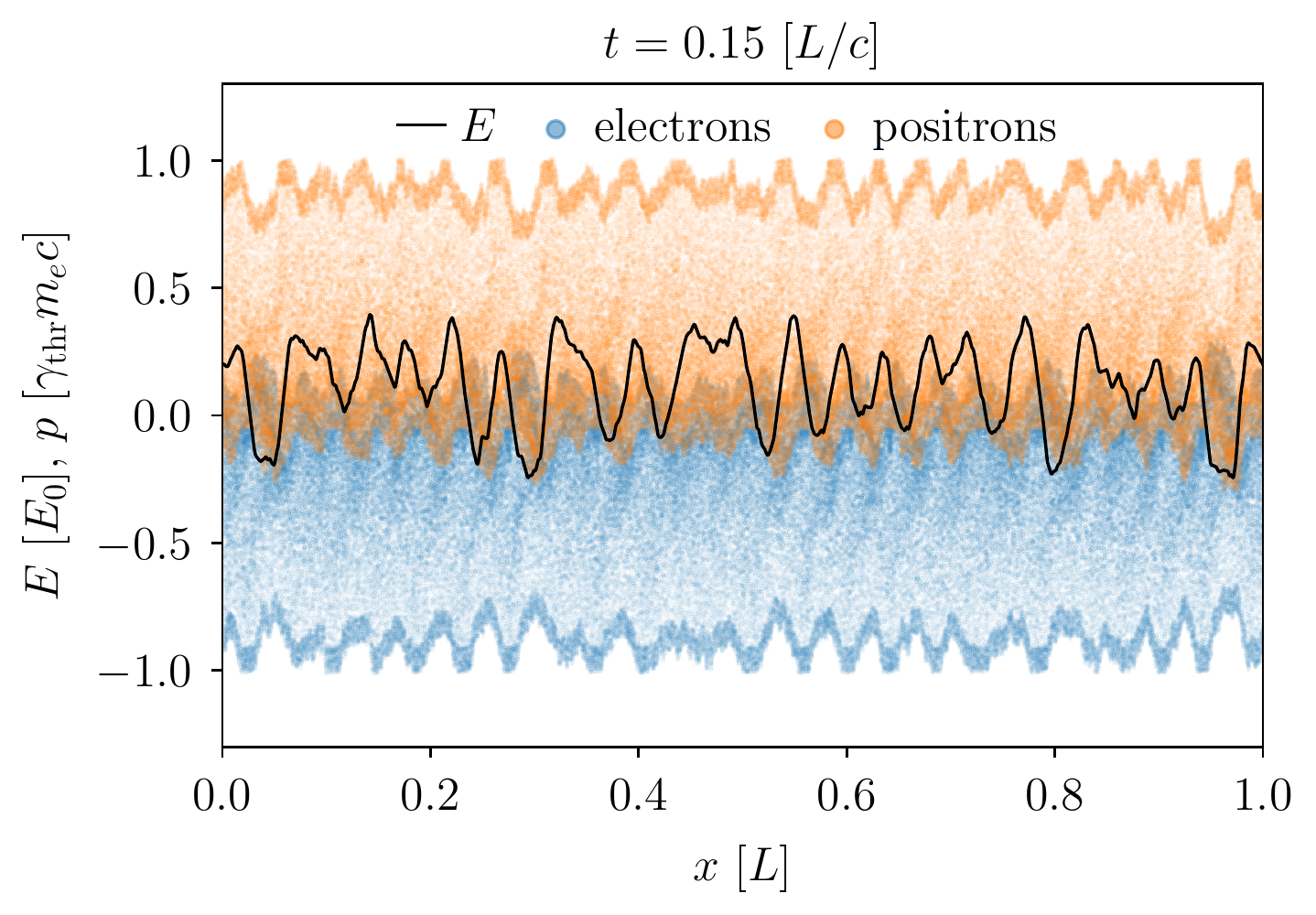}
\caption{\label{fig:const-e-phase} (Multimedia view) Phase space of electrons (blue dots) and positrons (orange dots) and electric field profile (black line). The static frame shows a time where the initial pair cascade has ceased, and the electric field oscillates inductively in time, \textit{i.e.}, supported by plasma currents. Perturbations of the electric field developed as a product of the instability of inductive waves are also visible. Reacceleration of electrons and positrons in these perturbations triggers secondary pair production bursts.}
\end{figure}

When the field is screened, the plasma current is maximum, which reverses the field, decelerating particles. The system enters a stage where the electric field is approximately constant in space and oscillates in time inductively, \textit{i.e.}, supported by reversals of the plasma current~\citep{levinson_2005}. The frequency of these oscillations is the relativistic plasma frequency~\citep{cruz_2021}, $\omega_0 \simeq \sqrt{ 4\pi e^2 n_\pm / \gamma_\mathrm{thr} m_e}$. This is illustrated in Fig.~\ref{fig:const-ene} for times $t \simeq (0.02 - 0.15) ~ L / c$. During this time, electrons and positrons periodically reverse their momentum without producing new pairs. With time, inductive oscillations become unstable~\citep{cruz_2021c}, and perturbations in the electric field accelerate a small fraction of particles beyond $\gamma_\mathrm{thr}$, producing new bursts of pair production. In Fig.~\ref{fig:const-e-phase} (Multimedia view), we show the phase space of electrons and positrons and the electric field profile to illustrate the development of these perturbations. The consecutive pair production bursts ultimately damp the average electric field, and energy is further transferred into particle kinetic energy, as shown in Fig.~\ref{fig:const-ene} for times $t > 0.15 ~ L / c$.

The system behaviour described above and illustrated in Figs.~\ref{fig:const-ene} and \ref{fig:const-e-phase} for $f = 0.1$ occurs similarly for other values of $f \ll 1$ and $\gamma_\mathrm{thr}$. In particular, we have performed simulations with $\gamma_\mathrm{thr} = 500 - 5000$ and $f = 10^{-3} - 0.1$, and the onset and saturation of the cascade, as well as the excitation and damping of inductive plasma waves, are qualitatively similar for all cases. All analytical estimates derived from the heuristic model considered here, such as the cascade growth rate, the time required for the electric field to be screened and the frequency of the inductive plasma waves, are consistent with simulation results.

\section{Cascade in a linear electric field}
\label{sec:linear}

We now consider a more complex and realistic profile for the electric field. In particular, we assume that the electric field has spatio-temporal dependence, $E = E(x, t)$. A good approximation of $E(x, t)$ can be obtained for vacuum gaps near the stellar surface. To describe the plasma near the surface of the neutron star, it is convenient to write Maxwell's equations in the frame that co-rotates with the pulsar. In this frame, Gauss's and Amp\`{e}re's laws are respectively written as~\citep{timokhin_2010}
\begin{subequations}
\begin{equation}
\frac{\partial E}{\partial x} = 4 \pi (\rho - \rho_\mathrm{GJ}) \ , 
\label{eq:gauss_corotating}
\end{equation}
\begin{equation}
\frac{\partial E}{\partial t} = - 4 \pi (j - j_\mathrm{m}) \ ,
\label{eq:ampere_corotating}
\end{equation}
\end{subequations}
where $\rho_\mathrm{GJ}$ is the Goldreich-Julian density~\citep{goldreich_julian_1969}, \textit{i.e.}, the equilibrium plasma density in the pulsar magnetosphere, and $j_\mathrm{m}$ is a background current that locally supports the twist of the magnetic field imposed at large scales by the magnetosphere.

\begin{figure}
\centering
\includegraphics[width=2.9in]{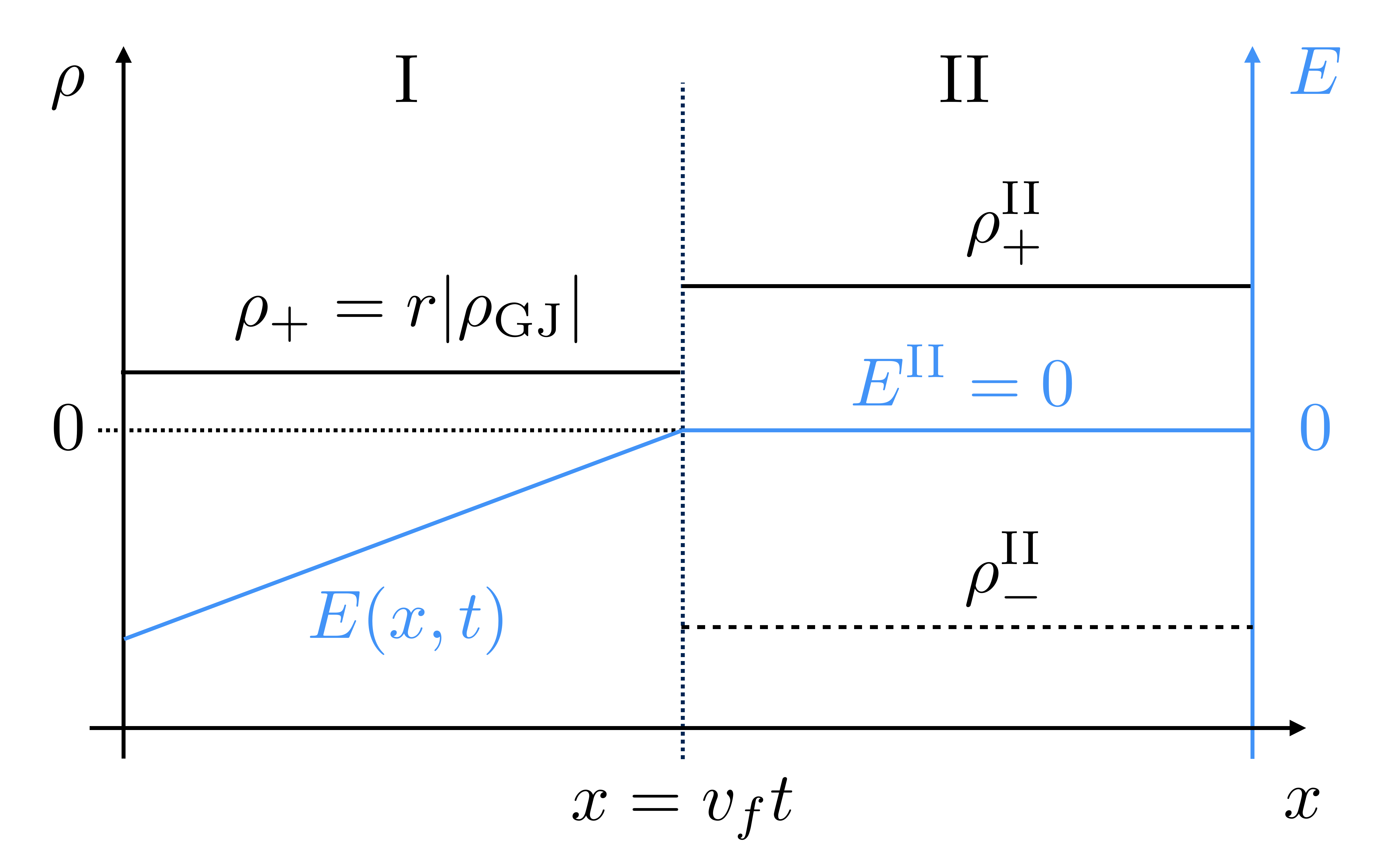}
\caption{\label{fig:electric_field_scheme} Schematic representation of the linear electric field profile considered in Sec.~\ref{sec:linear}. In the open gap (region I), a stream of positrons of charge density $\rho_+$ travels toward the stellar surface ($x = 0$). Because the total charge density in this region does not match $\rho_\mathrm{GJ}$, a linear electric field $E(x, t)$ given by Eq.~\eqref{eq:electric_field_linear} develops. In region II, the electric field is $E^\mathrm{II} = 0$ because the total charge and current densities match the equilibrium conditions, $\rho^\mathrm{II} = \rho_+^\mathrm{II} + \rho_-^\mathrm{II} = \rho_\mathrm{GJ}$ and $j^\mathrm{II} = j_\mathrm{m}$, respectively. The boundary between regions I and II moves with velocity $v_f > 0$.}
\end{figure}

When the plasma density $\rho$ and/or current density $j$ deviate from their equilibrium values, a vacuum gap develops. Simulations show~\citep{timokhin_2010, timokhin_2013, cruz_2021} that this precedes pair cascades, with only one species (electrons or positrons) of approximately constant density flowing through the gap toward the stellar surface. Without loss of generality, we assume here that positrons initially flow through the gap (which corresponds to the configuration found in gaps of pulsars where the magnetic and rotation axes are aligned, with $\rho_\mathrm{GJ} < 0$) at the speed of light with charge density $\rho_+ = r |\rho_\mathrm{GJ}|$, where $0 < r < 1$ is a ratio to be determined. In these conditions, it is possible to integrate Eqs.~\eqref{eq:gauss_corotating} and \eqref{eq:ampere_corotating} to show the electric field is
\begin{equation}
E(x, t) = 
\begin{cases}
4 \pi |\rho_\mathrm{GJ}| (1 + r) (x - v_f t) \ , x < v_f t \\
0 \ , x \ge v_f t \ ,
\end{cases}
\label{eq:electric_field_linear}
\end{equation}
where $v_f$ is the velocity at which the \textit{front} of the gap moves. This velocity can be related to $\alpha \equiv - j_\mathrm{m} / |\rho_\mathrm{GJ}| c$ and $r$ as $v_f / c = (\alpha - r) / (1 + r) \simeq r$. Previous analytical~\citep{beloborodov_2008} and numerical~\citep{timokhin_2013} works suggest that $\alpha < 0$ or $\alpha > 1$ for pair cascades to be produced in pulsar polar caps. In the following analysis, we consider that $1 < \alpha < 3$, such that $1/3 < v_f / c < 1$. The electric field profile in Eq.~\eqref{eq:electric_field_linear} is schematically illustrated in Fig.~\ref{fig:electric_field_scheme}.
% In our model, the maximum value of $\alpha$ is $3$, which corresponds to an equilibrium supported by an electron and positron flows of current densities $j_- = - 2 |\rho_\mathrm{GJ}| c$ and $j_+ = - |\rho_\mathrm{GJ}| c$, respectively. In these conditions $r = 1$, \textit{i.e.}, all positrons flow through the gap, and $v_f / c = 1$.

\subsection{Growth in time}
\label{sec:growth-time}

It is straightforward to note that the typical time scales $t_a$ and $t_p$ are no longer constant nor equal for electrons and positrons with the electric field in Eq.~\eqref{eq:electric_field_linear}. For $x < v_f t$, $E(x, t) < 0$ and thus any positron is accelerated toward the stellar surface ($x = 0$), and conversely any electron is accelerated toward the magnetosphere ($x > 0$). Therefore, in their paths, positrons/electrons experience increasing/decreasing values of the electric field. Thus, the acceleration and re-acceleration times of positrons and electrons are a function of the position $x_i$ and time $t_i$ in which they are created or enter the gap, $t_{a\pm} = t_{a\pm} (x_i, t_i)$ and $t_{p\pm} = t_{p\pm} (x_i, t_i)$. Integrating the equation of motion of electrons and positrons, it is possible to derive
\begin{align}
t_{a\pm} (x_i, t_i) &\simeq \frac{t_i \beta_f - x_i / c}{1 \pm \beta_f} - \nonumber \\
&- \sqrt{ \left( \frac{t_i \beta_f - x_i / c}{1 \pm \beta_f} \right)^2  \pm \frac{\gamma_\mathrm{thr} / (1 + r) \omega_{p, \mathrm{GJ}}^2}{1 \pm \beta_f} } \ ,
\label{eq:ta_linear}
\end{align}
where $\beta_f \equiv v_f / c$ and $\omega_{p, \mathrm{GJ}} = \sqrt{4 \pi e^2 n_\mathrm{GJ} / m_e}$ is the plasma frequency associated with a number density $n_\mathrm{GJ} = |\rho_\mathrm{GJ}| / e$. Under realistic conditions, the terms inside the square root in Eq.~\eqref{eq:ta_linear} can be comparable, so both need to be taken into account when computing $t_{a\pm}$. The reacceleration time $t_{p\pm}$ can be expressed by replacing $\gamma_\mathrm{thr} \to f \gamma_\mathrm{thr}$ in the last term of Eq.~\eqref{eq:ta_linear}.

Rigorously determining the evolution in time of the number of particles in the cascade with the time and space dependent $t_{a\pm}$ and $t_{p\pm}$ determined above is a challenging task. Moreover, with such time and space dependence, it is not straightforward that a pure exponential behavior similar to that observed for a constant electric field occurs. Here, we seek a solution for a scenario where $t_{a\pm}$ and $t_{p\pm}$ are slowly varying, such that one may have a local growth rate. In particular, we investigate the case of a single layer of electrons streaming through the open gap and pair producing over time. For such layer, new electrons always co-move with their primary particles, whereas positrons are immediately directed toward the stellar surface. We consider this layer to be arbitrarily thin, which is a good approximation since its thickness grows as $\Delta x \simeq t c / (f \gamma_\mathrm{thr})^2$, with $f \gamma_\mathrm{thr} \ll 1$. As mentioned above, we study in particular the case where $t_a (t)$ and $t_p (t)$ (we drop the $-$ subscript from here on for simplicity) are slowly evolving, which is possible for $v_f / c \gtrsim 0.7$ and $f \lesssim 0.05$. First, we assume that $t_p(t) \simeq f t_a(t)$ and therefore, since $f \ll 1$, $t_a(t) - 3 t_p(t) / 2 \simeq t_a(t)$. This assumption is valid as long as the layer is far from the front of the gap. In these conditions it is possible to rewrite Eqs.~\eqref{eq:cascade_delay} and \eqref{eq:cascade_deriv} as
\begin{subequations}
\begin{equation}
n_1 (t + t_a (t)) \simeq n_1(t) + n_2(t) \ ,
\label{eq:cascade_delay_linear}
\end{equation}

\begin{equation}
\frac{\mathrm{d} n_2(t)}{\mathrm{d} t} \simeq \frac{n_1(t)}{f t_a(t)} - \frac{n_2(t)}{t_a(t)} \ .
\label{eq:cascade_deriv_linear}
\end{equation}
\end{subequations}
We note that the first term on the right hand side of Eqs.~\eqref{eq:cascade_deriv} and \eqref{eq:cascade_deriv_linear} differs by a factor of $2$ because, in the case of a linear electric field, the growth in the number of electrons is caused only by pair production events triggered by primary electrons (and not positrons).

Given the results presented in Sec.~\ref{sec:uniform} and the rational above, it is reasonable to assume that the solution of this system can be written with the WKB approximation
\begin{equation}
n_{1, 2} (t) \propto \exp \left( \int_{t_a^*}^t \Gamma (t') \ \mathrm{d} t' \right) \ ,
\end{equation}
where $\Gamma(t')$ is the local (in time) growth rate of the cascade at time $t'$ and $t_a^*$ is the time in which $n_{1, 2}$ start to grow exponentially, which can be estimated as the time of creation of the second generation of electrons. Plugging $n_2(t)$ into Eq.~\eqref{eq:cascade_deriv_linear}, we can write
\begin{equation}
\frac{n_2(t)}{n_1(t)} = \frac{1}{f (\Gamma(t) t_a(t)+1)} \ .
\label{eq:cascade_ratio_linear}
\end{equation}
Assuming now that $\Gamma(t)$ varies slowly during a time $t_a(t)$, we use its Taylor expansion $\Gamma (t') = \Gamma (t) + (t' - t) \dot{\Gamma} (t)$ to obtain
\begin{align}
\int_{t}^{t+t_a(t)} \Gamma(t') \, \mathrm{d}t' & \simeq  \Gamma(t) t_a(t)  \left( 1 + \frac{\dot{\Gamma}(t) t_a(t)}{2 \Gamma(t)} \right) \nonumber \\
&\equiv \Gamma(t) t_a(t)  \left( 1+ \psi (t) \right) \ .
\end{align}
Plugging this result in Eq.~\eqref{eq:cascade_delay_linear} and using Eq.~\eqref{eq:cascade_ratio_linear}, we get
\begin{equation}
\exp \left[ \tilde{\Gamma}(t) t_a(t) \right] \simeq \frac{1+ \psi(t)}{f \tilde{\Gamma}(t) t_a(t) } \ ,
\label{eq:gamma_tilde}
\end{equation}
where $\tilde{\Gamma}(t) = \Gamma(t) ( 1+ \psi (t))$. The solution of Eq.~\eqref{eq:gamma_tilde} yields a correction to Eq.~\eqref{eq:gamma_uniform},
\begin{equation}
\Gamma(t) t_a(t)= \frac{1}{1+\psi(t)}   W\left (\frac{1+\psi(t)}{f} \right) \ .
\label{eq:gamma_linear}
\end{equation}
We note that this is not a fully closed form of $\Gamma (t)$, since $\psi$ is a function of $\Gamma$ and $\dot{\Gamma}$. Assuming that $t_a(t) \simeq t_a^* (1 + C t)$, where $C \ll 1$ is a constant, we can get $\psi$ to the lowest order when $\Gamma (t) t_a(t) \simeq W(1 / f)$,
\begin{equation}
\psi \simeq \frac{C t_a^*}{W(1 / f)} \ll 1 \ .
\end{equation}

We have performed a set of 1D PIC simulations that validate the analysis presented above. In these simulations, a single electron is initialized at the stellar surface ($x = 0$). This electron experiences an external electric field given by Eq.~\eqref{eq:electric_field_linear} with the time translation $t \to t - t_0$, where the time $t_0$ is estimated to be the time at which a positron stream from the magnetosphere would pair produce at the surface. We use in all simulations $\gamma_\mathrm{thr} = 1000$, and vary $v_f \in [0.5, 0.9]$ and $f \in [0.005, 0.1]$. The length of the simulation domain $L$ is chosen such that electron layer reaches the front of the gap at $x < L$. The grid resolution is $\Delta x / L = 0.001$ and the time step is chosen to resolve the smallest expected $t_p(t)$. Additionally, we set the initial electron density to be low enough such that particle motion is dominated by the externally imposed electric field at all times, \textit{i.e.}, the total electric field experienced by electrons is never significantly affected by the plasma current.

\begin{figure}[h]
\includegraphics[width=3.4in]{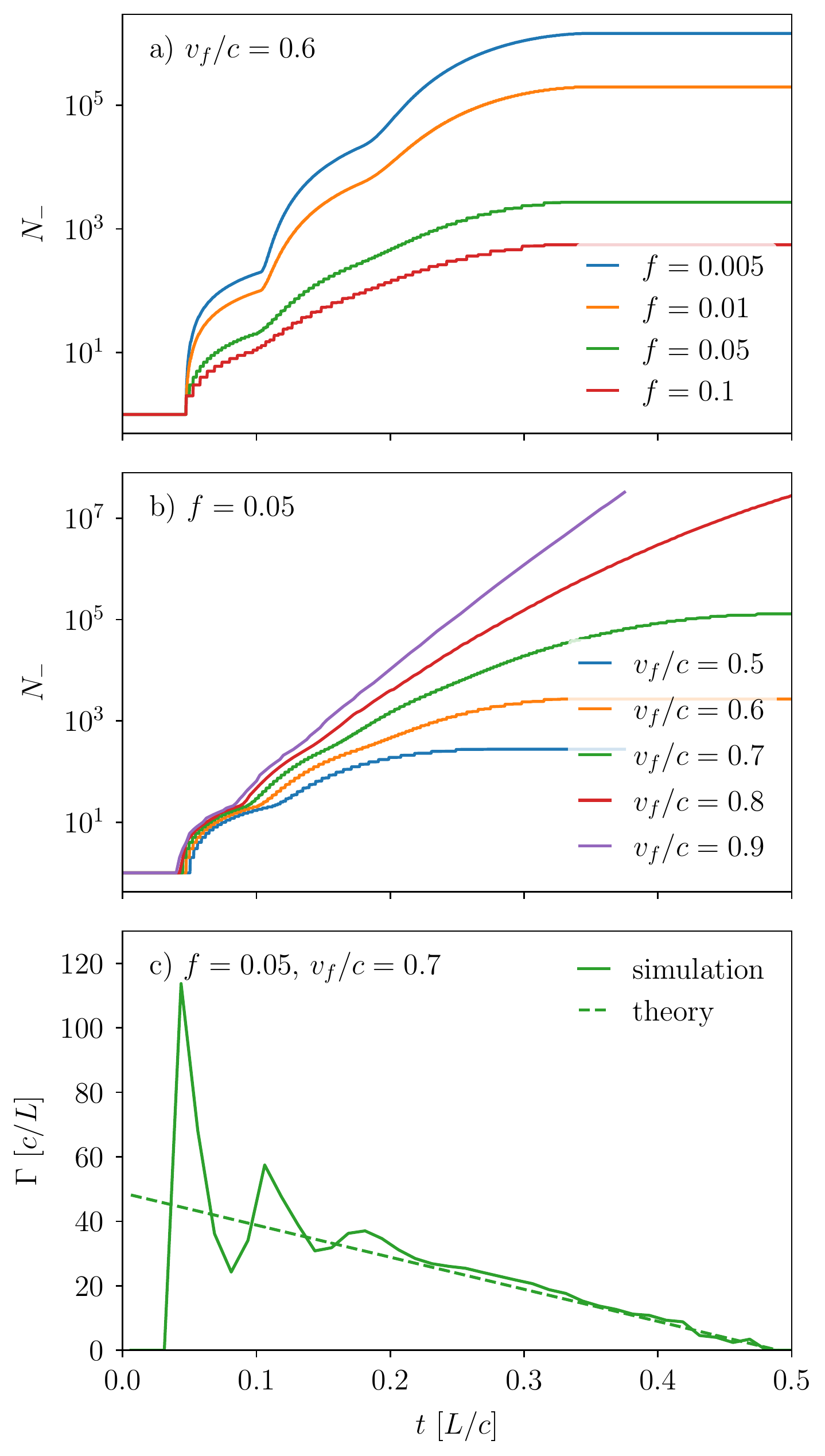}
\caption{\label{fig:single_electron_layer} Time evolution of the number of electrons $N_-$ in simulations of pair cascades developed in a linear electric field with a) fixed $v_f / c = 0.6$ and varying $f$ and b) fixed $f = 0.05$ and varying $v_f / c$. Panel c) shows the evolution of the local (in time) growth rate of $N_-$ for the simulation with $f = 0.05$ and $v_f / c = 0.7$ (represented with a green line in b)), together with the theoretical estimate given by Eq.~\eqref{eq:gamma_linear}.}
\end{figure}

In Fig.~\ref{fig:single_electron_layer}, we show the results of these simulations. Figures~\ref{fig:single_electron_layer} a) and b) show the evolution in time of the number of electrons $N_-$ for fixed $v_f = 0.6$ and varying $f$ and for fixed $f = 0.05$ and varying $v_f$, respectively. These results suggest that, as with a uniform background electric field, the cascade grows exponentially. It is also clear that, as before, the cascade grows faster for smaller $f$. In these simulations, the number of electrons grows until the electron layer catches the front of the gap, where $E = 0$ and pair production ceases. For the set of simulations shown in Fig.~\ref{fig:single_electron_layer} a), this happens at $t \simeq 0.3 ~ L/c$. In Fig.~\ref{fig:single_electron_layer} b) it is clear that this time increases with $v_f$, as intuitively expected. From Figs.~\ref{fig:single_electron_layer} a) and b), it is also clear that $N_-$ grows progressively slower with time, which is consistent with our hypothesis from the expected increase over time of $t_p(t)$ and $t_a(t)$. In Fig.~\ref{fig:single_electron_layer} c) we show the growth rate $\Gamma (t)$ for the simulation with $f = 0.05$ and $v_f/c = 0.7$. This growth rate was obtained by fitting exponential curves to $N_-(t)$ over short time intervals centered at $t$.  In Fig.~\ref{fig:single_electron_layer} c), we also plot the growth rate estimated with Eq.~\eqref{eq:gamma_linear}, showing that it approximates well the simulation results. A similar correspondence is obtained for all simulations with $v_f / c > 0.7$ and $f < 0.05$. We note that the peaks observed in Fig.~\ref{fig:single_electron_layer} c) correspond to times $t \lesssim 1 - 2 t_a$, in which the growth is not yet purely exponential, and thus fall out of the regime of validity of our model. This only approximately exponential phase is visible in Figs.~\ref{fig:single_electron_layer} a) and b) in the sudden periodic increases in $N_-$ at early times.

\subsection{Growth in space}
\label{sec:growth-space}

To extend our analysis from one to multiple electron layers, we have performed a second set of simulations where we externally impose the electric field in Eq.~\eqref{eq:electric_field_linear} (without time translation) and start with a uniform positron distribution. Positrons are accelerated as the gap opens, pair producing at $x \simeq 0$. Fresh electrons are then accelerated toward the magnetosphere, with each electron developing its own single-layered cascade. Since we start from a uniform positron distribution, a multi-layered cascade is generated in this case. However, all layers move at approximately the speed of light, so each layer behaves according to the model outlined above.

\begin{figure}[h]
\includegraphics[width=3.4in]{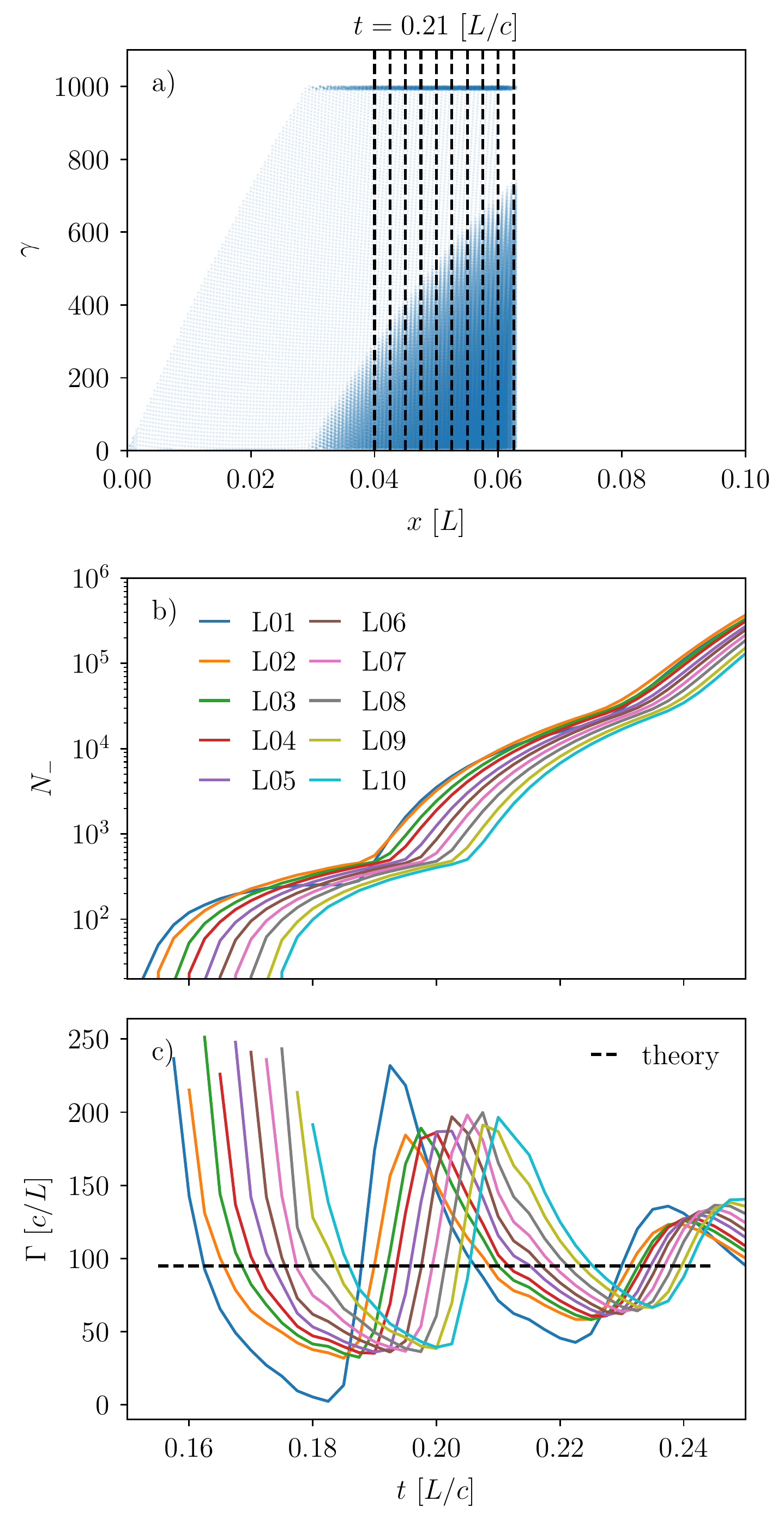}
\caption{\label{fig:multiple_electron_layer} a) Distribution in space and energy of electrons (blue dots) in simulation of a pair cascade with a linear electric field with parameters $f = 0.01$ and $v_f / c = 0.9$. b) Time evolution of the number of electrons in the layers identified by the boundaries drawn with vertical dashed lines in a). Layer numbers increase from the front to the back of the cascade (\textit{i.e.}, with decreasing $x$). c) Local (in time) growth rates of the curves in b), compared with the theoretical estimate given by Eq.~\eqref{eq:gamma_linear} with $\psi \simeq 0$ and $t_a(t) \simeq t_a^*$.}
\end{figure}

We illustrate the development of this multi-layered cascade in a simulation with $f = 0.01$ and $v_f / c = 0.9$ in Fig.~\ref{fig:multiple_electron_layer}. We follow the number of electrons in different layers of width $\Delta x_\mathrm{L} = 0.0025~L$ starting from the leading edge, or the \textit{front} of the cascade. In Fig.~\ref{fig:multiple_electron_layer} a), we present the particle distribution in space and energy, as well as a visual representation of the multiple layers considered. Figures~\ref{fig:multiple_electron_layer} b) and c) show the evolution in time of the number of electrons in each layer and the corresponding growth rate, respectively, confirming that each layer grows in an approximately exponential fashion. Furthermore, we see that the layers grow progressively faster from the front to the back of the cascade. This is a consequence of the electrons in the back layers experiencing a larger electric field at the time they are created (and thus throughout the development of their single-layered cascade). Moreover, in Fig.~\ref{fig:multiple_electron_layer} c) we also show that the growth rate obtained analytically (with Eq.~\eqref{eq:gamma_linear} with $\psi \simeq 0$ and $t_a(t) \simeq t_a^*$) fits reasonably well the average growth rate of the different layers.

\begin{figure}
\includegraphics[width=3.4in]{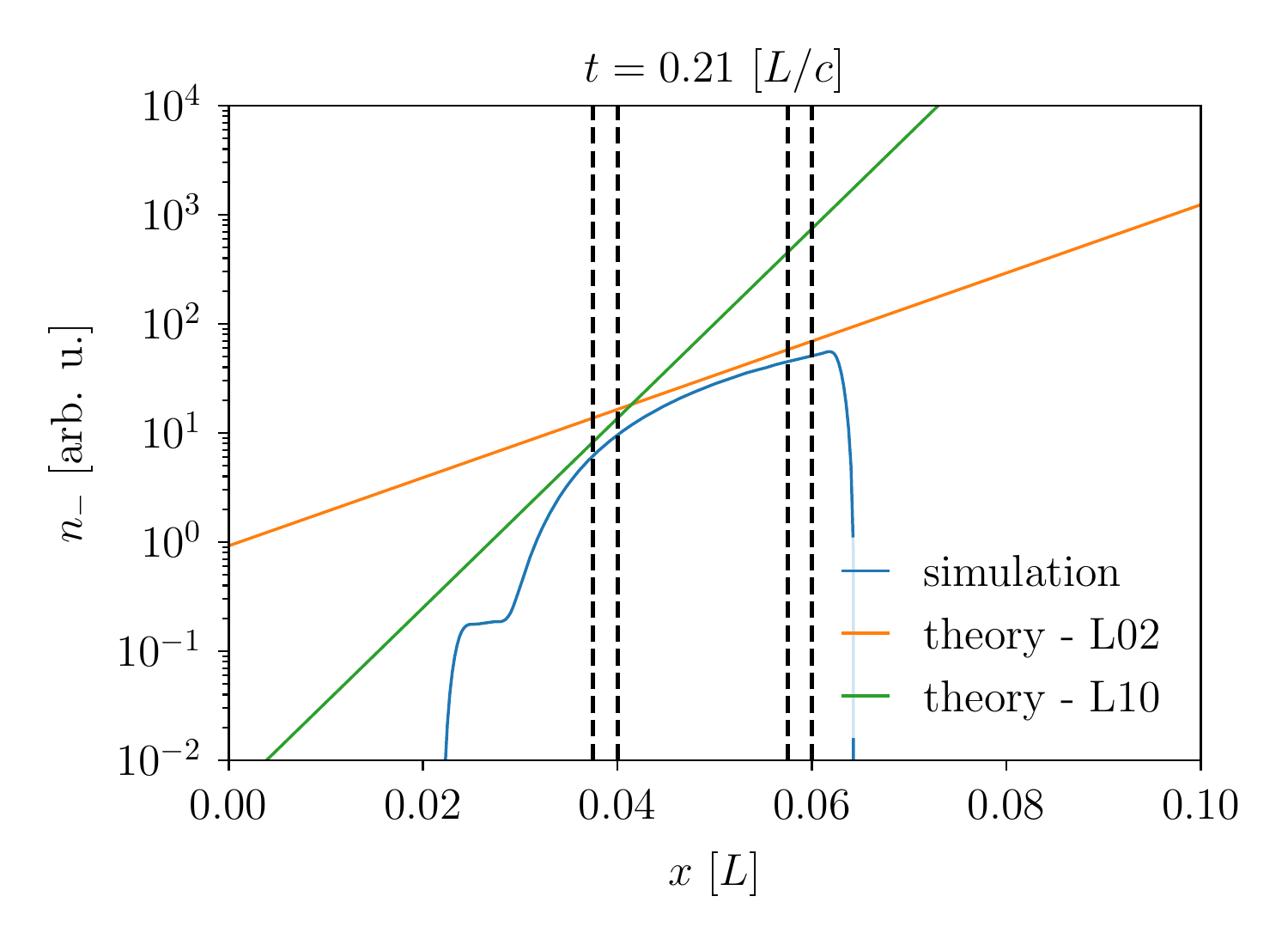}
\caption{\label{fig:spatial_growth} Electron density profile obtained in the same simulation illustrated in Fig.~\ref{fig:multiple_electron_layer} ($f = 0.01$ and $v_f / c = 0.9$). The density profile is well estimated by the theoretical estimate in Eq.~\eqref{eq:growth_space}, which is represented here with the slopes of the lines in orange and green for the layers L02 and L10 of the cascade, respectively. These layers are defined in the text and illustrated in Fig.~\ref{fig:multiple_electron_layer} a).}
\end{figure}

The varying growth rate identified above for the different layers naturally gives rise to a non uniform electron density spatial profile. This profile is shown in Fig.~\ref{fig:spatial_growth} for the same simulation and time shown in Fig.~\ref{fig:multiple_electron_layer} a). Despite the lower growth rate in time of the front layers, the density decreases from the front to the back of the cascade. This happens because layers at any distance $d$ behind the head are created with a temporal lag $d / c$, which for the time shown in Fig.~\ref{fig:spatial_growth} is not enough to flatten (or reverse) the electron density profile.

We can estimate the profile $n_- (x, t)$ based on the form of $\Gamma(t)$ previously determined. Assuming that the electric field varies slowly in space, we can write the density of the $k$th layer from the front as
\begin{equation}
n_{-, k} (t) \simeq n_{-, k+1} (t) \exp \left( \Gamma(t) \Delta x_\mathrm{L} / c \right) \ .
\end{equation}
Taking arbitrarily small $\Delta x_\mathrm{L}$, we can finally write
\begin{equation}
n_-(x, t) \propto \exp (\Gamma(t) ( t + x / c)) \ .
\label{eq:growth_space}
\end{equation}
This estimate holds for as long as the growth rate of individual layers is approximately the same, \textit{i.e.}, while the number of electrons in each layer is dominantly determined by the life time of that layer. This condition can be written as $t \dot{\Gamma} (t) / \Gamma (t) \ll 1$, which is valid in early times of the cascade. In Fig.~\ref{fig:spatial_growth}, we plot lines with slopes computed with Eq.~\eqref{eq:growth_space} for sample layers L02 and L10 which are shown to be in good agreement with the local shape of $n_- (x,t)$ obtained from the simulation.

The approximate profile in Eq.~\eqref{eq:growth_space} can finally be used to compute the time it takes to screen the electric field. Since the front of the cascade is the denser than the back, the field is first screened also at this position when $n_- \simeq - j_\mathrm{m} / e c$. Since $|j_\mathrm{m}| \simeq (1 - 2) |\rho_\mathrm{GJ}| c$ and, from previous simulations~\citep{timokhin_2010, timokhin_2013, cruz_2021}, $n_- (t)$ grows from an initial density $\simeq (0.01 - 0.1) |\rho_\mathrm{GJ}| / e$, then the screening time is $t_s \simeq 1 / \Gamma(t_a*)$. Given that $\Gamma(t_a^*) \simeq W(1 / f) t_a(t_a^*)$ (from Eq.~\eqref{eq:gamma_linear}) and that $W (1 / f) \simeq 3 - 5$ for $f \simeq 10^{-6} - 10^{-3}$, we can finally estimate $t_s \simeq t_a(t_a^*)$. For realistic pulsars, $t_s$ varies between $10^{-9}$ and $10^{-6}$~s with increasing rotation period, a result in good agreement with previous theoretical and numerical predictions~\citep{timokhin_2015}.

\section{Conclusions}
\label{sec:conclusions}

Heuristic models of pair cascades represent an important tool to develop a theoretical understanding of the rich interplay between QED and plasma kinetic effects. In this work, we have studied this interplay using an heuristic model in which electrons and positrons produce new pairs whenever they are accelerated beyond a threshold energy. In particular, we have compared the development of cascades in constant and linear background electric field profiles. For a uniform electric field, we have shown that the number of particles in the cascade grows exponentially, and compared an analytical estimate of its growth rate with simulation results. 

We have then shown that cascades also grow exponentially in a linear electric field, but present a key distinctive property: the growth rate decreases with time due to the decreasing electric field that the cascade seed particles experience. This ultimately results in a spatially dependent growth rate for the density of electron clouds generated in cascades near the surface of neutron stars. Specifically, the density of the front of these clouds is expected to grow slower than their back. However, since the front of the clouds is generated earlier, the electron density decreases from the front to the back. These results are consistent with previous 1D PIC simulations of pair cascades including pair production from first principles and in self-consistently developed (linear) electric fields~\citep{timokhin_2010, timokhin_2013}. We have derived analytical estimates for the spatio-temporal growth of the cascade and validated them against 1D PIC simulations including heuristic pair production. Finally, we have estimated the time at which the plasma generated in the cascade first screens the gap electric field. An analysis of the properties of inductive waves self-consistently generated as a result of pair cascades in a linear electric field is left as future work.

% More complex models
The insights developed with the analytical models presented in this work can be applied to interpret more complex, full QED simulations of pair cascades. Yet, their applicability is limited to the regime of validity of the heuristic model considered in this work (negligible photon mean free path). Extensions of this model, and consequently of the theory presented here, are next steps on the study of pair cascades in compact objects. In particular, including a soft energy threshold for photon emission and a (constant or spatially dependent) finite photon mean free path would further approximate these models to the \textit{ab initio} description of pair cascades.

\begin{acknowledgments}
This work was supported by the European Research Council (ERC-2015-AdG Grant 695088) and FCT (Portugal) (grant PD/BD/114307/2016) in the framework of the Advanced Program in Plasma Science and Engineering (APPLAuSE, FCT grant PD/00505/2012). % We acknowledge PRACE for granting access to MareNostrum, Barcelona Supercomputing Center (Spain), where the simulations presented in this work were performed.
\end{acknowledgments}

\bibliography{aiptemplate}

\begin{thebibliography}{39}%
\makeatletter
\providecommand \@ifxundefined [1]{%
 \@ifx{#1\undefined}
}%
\providecommand \@ifnum [1]{%
 \ifnum #1\expandafter \@firstoftwo
 \else \expandafter \@secondoftwo
 \fi
}%
\providecommand \@ifx [1]{%
 \ifx #1\expandafter \@firstoftwo
 \else \expandafter \@secondoftwo
 \fi
}%
\providecommand \natexlab [1]{#1}%
\providecommand \enquote  [1]{``#1''}%
\providecommand \bibnamefont  [1]{#1}%
\providecommand \bibfnamefont [1]{#1}%
\providecommand \citenamefont [1]{#1}%
\providecommand \href@noop [0]{\@secondoftwo}%
\providecommand \href [0]{\begingroup \@sanitize@url \@href}%
\providecommand \@href[1]{\@@startlink{#1}\@@href}%
\providecommand \@@href[1]{\endgroup#1\@@endlink}%
\providecommand \@sanitize@url [0]{\catcode `\\12\catcode `\$12\catcode
  `\&12\catcode `\#12\catcode `\^12\catcode `\_12\catcode `\%12\relax}%
\providecommand \@@startlink[1]{}%
\providecommand \@@endlink[0]{}%
\providecommand \url  [0]{\begingroup\@sanitize@url \@url }%
\providecommand \@url [1]{\endgroup\@href {#1}{\urlprefix }}%
\providecommand \urlprefix  [0]{URL }%
\providecommand \Eprint [0]{\href }%
\providecommand \doibase [0]{http://dx.doi.org/}%
\providecommand \selectlanguage [0]{\@gobble}%
\providecommand \bibinfo  [0]{\@secondoftwo}%
\providecommand \bibfield  [0]{\@secondoftwo}%
\providecommand \translation [1]{[#1]}%
\providecommand \BibitemOpen [0]{}%
\providecommand \bibitemStop [0]{}%
\providecommand \bibitemNoStop [0]{.\EOS\space}%
\providecommand \EOS [0]{\spacefactor3000\relax}%
\providecommand \BibitemShut  [1]{\csname bibitem#1\endcsname}%
\let\auto@bib@innerbib\@empty
%</preamble>
\bibitem [{\citenamefont {Michel}(1982)}]{michel_1982}%
  \BibitemOpen
  \bibfield  {author} {\bibinfo {author} {\bibfnamefont {F.~C.}\ \bibnamefont
  {Michel}},\ }\bibfield  {title} {\enquote {\bibinfo {title} {Theory of pulsar
  magnetospheres},}\ }\href {\doibase 10.1103/RevModPhys.54.1} {\bibfield
  {journal} {\bibinfo  {journal} {Rev. Mod. Phys.}\ }\textbf {\bibinfo {volume}
  {54}},\ \bibinfo {pages} {1--66} (\bibinfo {year} {1982})}\BibitemShut
  {NoStop}%
\bibitem [{\citenamefont {Pétri}(2016)}]{petri_2016}%
  \BibitemOpen
  \bibfield  {author} {\bibinfo {author} {\bibfnamefont {J.}~\bibnamefont
  {Pétri}},\ }\bibfield  {title} {\enquote {\bibinfo {title} {Theory of pulsar
  magnetosphere and wind},}\ }\href {\doibase 10.1017/S0022377816000763}
  {\bibfield  {journal} {\bibinfo  {journal} {J. Plasma Phys.}\ }\textbf
  {\bibinfo {volume} {82}},\ \bibinfo {pages} {635820502} (\bibinfo {year}
  {2016})}\BibitemShut {NoStop}%
\bibitem [{\citenamefont {Cerutti}\ and\ \citenamefont
  {Beloborodov}(2017)}]{cerutti_2017}%
  \BibitemOpen
  \bibfield  {author} {\bibinfo {author} {\bibfnamefont {B.}~\bibnamefont
  {Cerutti}}\ and\ \bibinfo {author} {\bibfnamefont {A.~M.}\ \bibnamefont
  {Beloborodov}},\ }\bibfield  {title} {\enquote {\bibinfo {title}
  {Electrodynamics of pulsar magnetospheres},}\ }\href {\doibase
  10.1007/s11214-016-0315-7} {\bibfield  {journal} {\bibinfo  {journal} {Space
  Sci. Rev.}\ }\textbf {\bibinfo {volume} {2017}},\ \bibinfo {pages} {111--136}
  (\bibinfo {year} {2017})}\BibitemShut {NoStop}%
\bibitem [{\citenamefont {{Coroniti}}(1990)}]{coroniti_1990}%
  \BibitemOpen
  \bibfield  {author} {\bibinfo {author} {\bibfnamefont {F.~V.}\ \bibnamefont
  {{Coroniti}}},\ }\bibfield  {title} {\enquote {\bibinfo {title}
  {{Magnetically Striped Relativistic Magnetohydrodynamic Winds: The Crab
  Nebula Revisited}},}\ }\href {\doibase 10.1086/168340} {\bibfield  {journal}
  {\bibinfo  {journal} {Astrophys. J.}\ }\textbf {\bibinfo {volume} {349}},\
  \bibinfo {pages} {538} (\bibinfo {year} {1990})}\BibitemShut {NoStop}%
\bibitem [{\citenamefont {{Lyubarskii}}(1996)}]{lyubarskii_1996}%
  \BibitemOpen
  \bibfield  {author} {\bibinfo {author} {\bibfnamefont {Y.~E.}\ \bibnamefont
  {{Lyubarskii}}},\ }\bibfield  {title} {\enquote {\bibinfo {title} {{A model
  for the energetic emission from pulsars.}}}\ }\href@noop {} {\bibfield
  {journal} {\bibinfo  {journal} {Astron. Astrophys.}\ }\textbf {\bibinfo
  {volume} {311}},\ \bibinfo {pages} {172--178} (\bibinfo {year}
  {1996})}\BibitemShut {NoStop}%
\bibitem [{\citenamefont {Sturrock}(1971)}]{sturrock_1971}%
  \BibitemOpen
  \bibfield  {author} {\bibinfo {author} {\bibfnamefont {P.~A.}\ \bibnamefont
  {Sturrock}},\ }\href {\doibase 10.1086/150865} {\bibfield  {journal}
  {\bibinfo  {journal} {Astrophys. J.}\ }\textbf {\bibinfo {volume} {164}},\
  \bibinfo {pages} {529--556} (\bibinfo {year} {1971})}\BibitemShut {NoStop}%
\bibitem [{\citenamefont {{Ruderman}}\ and\ \citenamefont
  {{Sutherland}}(1975)}]{ruderman_sutherland_1975}%
  \BibitemOpen
  \bibfield  {author} {\bibinfo {author} {\bibfnamefont {M.~A.}\ \bibnamefont
  {{Ruderman}}}\ and\ \bibinfo {author} {\bibfnamefont {P.~G.}\ \bibnamefont
  {{Sutherland}}},\ }\href {\doibase 10.1086/153393} {\bibfield  {journal}
  {\bibinfo  {journal} {Astrophys. J.}\ }\textbf {\bibinfo {volume} {196}},\
  \bibinfo {pages} {51--72} (\bibinfo {year} {1975})}\BibitemShut {NoStop}%
\bibitem [{\citenamefont {Erber}(1966)}]{erber_1966}%
  \BibitemOpen
  \bibfield  {author} {\bibinfo {author} {\bibfnamefont {T.}~\bibnamefont
  {Erber}},\ }\bibfield  {title} {\enquote {\bibinfo {title} {High-energy
  electromagnetic conversion processes in intense magnetic fields},}\ }\href
  {\doibase 10.1103/RevModPhys.38.626} {\bibfield  {journal} {\bibinfo
  {journal} {Rev. Mod. Phys.}\ }\textbf {\bibinfo {volume} {38}},\ \bibinfo
  {pages} {626--659} (\bibinfo {year} {1966})}\BibitemShut {NoStop}%
\bibitem [{\citenamefont {Ritus}(1985)}]{ritus_1985}%
  \BibitemOpen
  \bibfield  {author} {\bibinfo {author} {\bibfnamefont {V.~I.}\ \bibnamefont
  {Ritus}},\ }\href {\doibase 10.1007/BF01120220} {\bibfield  {journal}
  {\bibinfo  {journal} {J. Sov. Laser Res.}\ }\textbf {\bibinfo {volume} {6}},\
  \bibinfo {pages} {497--617} (\bibinfo {year} {1985})}\BibitemShut {NoStop}%
\bibitem [{\citenamefont {Arons}(1983)}]{arons_1983}%
  \BibitemOpen
  \bibfield  {author} {\bibinfo {author} {\bibfnamefont {J.}~\bibnamefont
  {Arons}},\ }\bibfield  {title} {\enquote {\bibinfo {title} {Pair creation
  above pulsar polar caps - geometrical structure and energetics of slot
  gaps},}\ }\href {\doibase 10.1086/160771} {\bibfield  {journal} {\bibinfo
  {journal} {Astrophys. J.}\ }\textbf {\bibinfo {volume} {266}},\ \bibinfo
  {pages} {215--241} (\bibinfo {year} {1983})}\BibitemShut {NoStop}%
\bibitem [{\citenamefont {Cheng}, \citenamefont {Ho},\ and\ \citenamefont
  {Ruderman}(1986)}]{cheng_1986}%
  \BibitemOpen
  \bibfield  {author} {\bibinfo {author} {\bibfnamefont {K.~S.}\ \bibnamefont
  {Cheng}}, \bibinfo {author} {\bibfnamefont {C.}~\bibnamefont {Ho}}, \ and\
  \bibinfo {author} {\bibfnamefont {M.}~\bibnamefont {Ruderman}},\ }\bibfield
  {title} {\enquote {\bibinfo {title} {Energetic radiation from rapidly
  spinning pulsars. {I. Outer} magnetosphere gaps},}\ }\href {\doibase
  10.1086/163829} {\bibfield  {journal} {\bibinfo  {journal} {Astrophys. J.}\
  }\textbf {\bibinfo {volume} {300}},\ \bibinfo {pages} {500} (\bibinfo {year}
  {1986})}\BibitemShut {NoStop}%
\bibitem [{\citenamefont {Timokhin}(2010)}]{timokhin_2010}%
  \BibitemOpen
  \bibfield  {author} {\bibinfo {author} {\bibfnamefont {A.~N.}\ \bibnamefont
  {Timokhin}},\ }\bibfield  {title} {\enquote {\bibinfo {title} {Time-dependent
  pair cascades in magnetospheres of neutron stars - {I}. {Dynamics} of the
  polar cap cascade with no particle supply from the neutron star surface},}\
  }\href {\doibase 10.1111/j.1365-2966.2010.17286.x} {\bibfield  {journal}
  {\bibinfo  {journal} {Mon. Not. R. Astron. Soc.}\ }\textbf {\bibinfo {volume}
  {408}},\ \bibinfo {pages} {2092--2114} (\bibinfo {year} {2010})}\BibitemShut
  {NoStop}%
\bibitem [{\citenamefont {Levinson}\ \emph {et~al.}(2005)\citenamefont
  {Levinson}, \citenamefont {Melrose}, \citenamefont {Judge},\ and\
  \citenamefont {Luo}}]{levinson_2005}%
  \BibitemOpen
  \bibfield  {author} {\bibinfo {author} {\bibfnamefont {A.}~\bibnamefont
  {Levinson}}, \bibinfo {author} {\bibfnamefont {D.}~\bibnamefont {Melrose}},
  \bibinfo {author} {\bibfnamefont {A.}~\bibnamefont {Judge}}, \ and\ \bibinfo
  {author} {\bibfnamefont {Q.}~\bibnamefont {Luo}},\ }\href {\doibase
  10.1086/432498} {\bibfield  {journal} {\bibinfo  {journal} {Astrophys. J.}\
  }\textbf {\bibinfo {volume} {631}},\ \bibinfo {pages} {456--465} (\bibinfo
  {year} {2005})}\BibitemShut {NoStop}%
\bibitem [{\citenamefont {Timokhin}\ and\ \citenamefont
  {Arons}(2012)}]{timokhin_2013}%
  \BibitemOpen
  \bibfield  {author} {\bibinfo {author} {\bibfnamefont {A.~N.}\ \bibnamefont
  {Timokhin}}\ and\ \bibinfo {author} {\bibfnamefont {J.}~\bibnamefont
  {Arons}},\ }\bibfield  {title} {\enquote {\bibinfo {title} {Current flow and
  pair creation at low altitude in rotation-powered pulsars' force-free
  magnetospheres: space charge limited flow},}\ }\href {\doibase
  10.1093/mnras/sts298} {\bibfield  {journal} {\bibinfo  {journal} {Mon. Not.
  R. Astron. Soc.}\ }\textbf {\bibinfo {volume} {429}},\ \bibinfo {pages}
  {20--54} (\bibinfo {year} {2012})}\BibitemShut {NoStop}%
\bibitem [{\citenamefont {Timokhin}\ and\ \citenamefont
  {Harding}(2015)}]{timokhin_2015}%
  \BibitemOpen
  \bibfield  {author} {\bibinfo {author} {\bibfnamefont {A.~N.}\ \bibnamefont
  {Timokhin}}\ and\ \bibinfo {author} {\bibfnamefont {A.~K.}\ \bibnamefont
  {Harding}},\ }\bibfield  {title} {\enquote {\bibinfo {title} {On the polar
  cap cascade pair multiplicity of young pulsars},}\ }\href {\doibase
  10.1088/0004-637x/810/2/144} {\bibfield  {journal} {\bibinfo  {journal}
  {Astrophys. J.}\ }\textbf {\bibinfo {volume} {810}},\ \bibinfo {pages} {144}
  (\bibinfo {year} {2015})}\BibitemShut {NoStop}%
\bibitem [{\citenamefont {Philippov}, \citenamefont {Timokhin},\ and\
  \citenamefont {Spitkovsky}(2020)}]{philippov_2020}%
  \BibitemOpen
  \bibfield  {author} {\bibinfo {author} {\bibfnamefont {A.}~\bibnamefont
  {Philippov}}, \bibinfo {author} {\bibfnamefont {A.}~\bibnamefont {Timokhin}},
  \ and\ \bibinfo {author} {\bibfnamefont {A.}~\bibnamefont {Spitkovsky}},\
  }\href {\doibase 10.1103/PhysRevLett.124.245101} {\bibfield  {journal}
  {\bibinfo  {journal} {Phys. Rev. Lett.}\ }\textbf {\bibinfo {volume} {124}},\
  \bibinfo {pages} {245101} (\bibinfo {year} {2020})}\BibitemShut {NoStop}%
\bibitem [{\citenamefont {Cruz}\ \emph {et~al.}(2021)\citenamefont {Cruz},
  \citenamefont {Grismayer}, \citenamefont {Chen}, \citenamefont {Spitkovsky},\
  and\ \citenamefont {Silva}}]{cruz_2021b}%
  \BibitemOpen
  \bibfield  {author} {\bibinfo {author} {\bibfnamefont {F.}~\bibnamefont
  {Cruz}}, \bibinfo {author} {\bibfnamefont {T.}~\bibnamefont {Grismayer}},
  \bibinfo {author} {\bibfnamefont {A.~Y.}\ \bibnamefont {Chen}}, \bibinfo
  {author} {\bibfnamefont {A.}~\bibnamefont {Spitkovsky}}, \ and\ \bibinfo
  {author} {\bibfnamefont {L.~O.}\ \bibnamefont {Silva}},\ }\bibfield  {title}
  {\enquote {\bibinfo {title} {Coherent emission from qed cascades in pulsar
  polar caps},}\ }\href {\doibase 10.3847/2041-8213/ac2157} {\bibfield
  {journal} {\bibinfo  {journal} {Astrophys. J. Lett.}\ }\textbf {\bibinfo
  {volume} {919}},\ \bibinfo {pages} {L4} (\bibinfo {year} {2021})}\BibitemShut
  {NoStop}%
\bibitem [{\citenamefont {{Philippov}}\ and\ \citenamefont
  {{Spitkovsky}}(2014)}]{philippov_2014}%
  \BibitemOpen
  \bibfield  {author} {\bibinfo {author} {\bibfnamefont {A.~A.}\ \bibnamefont
  {{Philippov}}}\ and\ \bibinfo {author} {\bibfnamefont {A.}~\bibnamefont
  {{Spitkovsky}}},\ }\bibfield  {title} {\enquote {\bibinfo {title}
  {{\textit{Ab initio} pulsar magnetosphere: three-dimensional particle-in-cell
  simulations of axisymmetric Pulsars}},}\ }\href {\doibase
  10.1088/2041-8205/785/2/L33} {\bibfield  {journal} {\bibinfo  {journal}
  {Astrophys. J. Lett.}\ }\textbf {\bibinfo {volume} {785}},\ \bibinfo {pages}
  {L33} (\bibinfo {year} {2014})}\BibitemShut {NoStop}%
\bibitem [{\citenamefont {Belyaev}(2015)}]{belyaev_2015}%
  \BibitemOpen
  \bibfield  {author} {\bibinfo {author} {\bibfnamefont {M.~A.}\ \bibnamefont
  {Belyaev}},\ }\bibfield  {title} {\enquote {\bibinfo {title} {Dissipation,
  energy transfer, and spin-down luminosity in {2.5D} {PIC} simulations of the
  pulsar magnetosphere},}\ }\href {\doibase 10.1093/mnras/stv468} {\bibfield
  {journal} {\bibinfo  {journal} {Mon. Not. R. Astron. Soc.}\ }\textbf
  {\bibinfo {volume} {449}},\ \bibinfo {pages} {2759--2767} (\bibinfo {year}
  {2015})}\BibitemShut {NoStop}%
\bibitem [{\citenamefont {{Cerutti}}\ \emph {et~al.}(2015)\citenamefont
  {{Cerutti}}, \citenamefont {{Philippov}}, \citenamefont {{Parfrey}},\ and\
  \citenamefont {{Spitkovsky}}}]{cerutti_2015}%
  \BibitemOpen
  \bibfield  {author} {\bibinfo {author} {\bibfnamefont {B.}~\bibnamefont
  {{Cerutti}}}, \bibinfo {author} {\bibfnamefont {A.~A.}\ \bibnamefont
  {{Philippov}}}, \bibinfo {author} {\bibfnamefont {K.}~\bibnamefont
  {{Parfrey}}}, \ and\ \bibinfo {author} {\bibfnamefont {A.}~\bibnamefont
  {{Spitkovsky}}},\ }\bibfield  {title} {\enquote {\bibinfo {title} {{Particle
  acceleration in axisymmetric pulsar current sheets}},}\ }\href {\doibase
  10.1093/mnras/stv042} {\bibfield  {journal} {\bibinfo  {journal} {Mon. Not.
  R. Astron. Soc.}\ }\textbf {\bibinfo {volume} {448}},\ \bibinfo {pages}
  {606--619} (\bibinfo {year} {2015})}\BibitemShut {NoStop}%
\bibitem [{\citenamefont {Kalapotharakos}\ \emph {et~al.}(2018)\citenamefont
  {Kalapotharakos}, \citenamefont {Brambilla}, \citenamefont {Timokhin},
  \citenamefont {Harding},\ and\ \citenamefont
  {Kazanas}}]{kalapotharakos_2018}%
  \BibitemOpen
  \bibfield  {author} {\bibinfo {author} {\bibfnamefont {C.}~\bibnamefont
  {Kalapotharakos}}, \bibinfo {author} {\bibfnamefont {G.}~\bibnamefont
  {Brambilla}}, \bibinfo {author} {\bibfnamefont {A.}~\bibnamefont {Timokhin}},
  \bibinfo {author} {\bibfnamefont {A.~K.}\ \bibnamefont {Harding}}, \ and\
  \bibinfo {author} {\bibfnamefont {D.}~\bibnamefont {Kazanas}},\ }\bibfield
  {title} {\enquote {\bibinfo {title} {Three-dimensional kinetic pulsar
  magnetosphere models: Connecting to gamma-ray observations},}\ }\href
  {\doibase 10.3847/1538-4357/aab550} {\bibfield  {journal} {\bibinfo
  {journal} {Astrophys. J.}\ }\textbf {\bibinfo {volume} {857}},\ \bibinfo
  {pages} {44} (\bibinfo {year} {2018})}\BibitemShut {NoStop}%
\bibitem [{\citenamefont {Brambilla}\ \emph {et~al.}(2018)\citenamefont
  {Brambilla}, \citenamefont {Kalapotharakos}, \citenamefont {Timokhin},
  \citenamefont {Harding},\ and\ \citenamefont {Kazanas}}]{brambilla_2018}%
  \BibitemOpen
  \bibfield  {author} {\bibinfo {author} {\bibfnamefont {G.}~\bibnamefont
  {Brambilla}}, \bibinfo {author} {\bibfnamefont {C.}~\bibnamefont
  {Kalapotharakos}}, \bibinfo {author} {\bibfnamefont {A.~N.}\ \bibnamefont
  {Timokhin}}, \bibinfo {author} {\bibfnamefont {A.~K.}\ \bibnamefont
  {Harding}}, \ and\ \bibinfo {author} {\bibfnamefont {D.}~\bibnamefont
  {Kazanas}},\ }\bibfield  {title} {\enquote {\bibinfo {title}
  {Electron{\textendash}positron pair flow and current composition in the
  pulsar magnetosphere},}\ }\href {\doibase 10.3847/1538-4357/aab3e1}
  {\bibfield  {journal} {\bibinfo  {journal} {Astrophys. J.}\ }\textbf
  {\bibinfo {volume} {858}},\ \bibinfo {pages} {81} (\bibinfo {year}
  {2018})}\BibitemShut {NoStop}%
\bibitem [{\citenamefont {Cruz}, \citenamefont {Grismayer},\ and\ \citenamefont
  {Silva}(2021{\natexlab{a}})}]{cruz_2021}%
  \BibitemOpen
  \bibfield  {author} {\bibinfo {author} {\bibfnamefont {F.}~\bibnamefont
  {Cruz}}, \bibinfo {author} {\bibfnamefont {T.}~\bibnamefont {Grismayer}}, \
  and\ \bibinfo {author} {\bibfnamefont {L.~O.}\ \bibnamefont {Silva}},\
  }\bibfield  {title} {\enquote {\bibinfo {title} {Kinetic model of
  large-amplitude oscillations in neutron star pair cascades},}\ }\href
  {\doibase 10.3847/1538-4357/abd2c0} {\bibfield  {journal} {\bibinfo
  {journal} {Astrophys. J.}\ }\textbf {\bibinfo {volume} {908}},\ \bibinfo
  {pages} {149} (\bibinfo {year} {2021}{\natexlab{a}})}\BibitemShut {NoStop}%
\bibitem [{\citenamefont {Chen}\ and\ \citenamefont
  {Beloborodov}(2014)}]{chen_2014}%
  \BibitemOpen
  \bibfield  {author} {\bibinfo {author} {\bibfnamefont {A.~Y.}\ \bibnamefont
  {Chen}}\ and\ \bibinfo {author} {\bibfnamefont {A.~M.}\ \bibnamefont
  {Beloborodov}},\ }\bibfield  {title} {\enquote {\bibinfo {title}
  {Electrodynamics of axisymmetric pulsar magnetosphere with electron-positron
  discharge: a numerical experiment},}\ }\href {\doibase
  10.1088/2041-8205/795/1/l22} {\bibfield  {journal} {\bibinfo  {journal}
  {Astrophys. J. Lett.}\ }\textbf {\bibinfo {volume} {795}},\ \bibinfo {pages}
  {L22} (\bibinfo {year} {2014})}\BibitemShut {NoStop}%
\bibitem [{\citenamefont {Philippov}, \citenamefont {Spitkovsky},\ and\
  \citenamefont {Cerutti}(2015)}]{philippov_2015a}%
  \BibitemOpen
  \bibfield  {author} {\bibinfo {author} {\bibfnamefont {A.~A.}\ \bibnamefont
  {Philippov}}, \bibinfo {author} {\bibfnamefont {A.}~\bibnamefont
  {Spitkovsky}}, \ and\ \bibinfo {author} {\bibfnamefont {B.}~\bibnamefont
  {Cerutti}},\ }\bibfield  {title} {\enquote {\bibinfo {title} {\textit{Ab
  initio} pulsar magnetosphere: three-dimensional particle-in-cell simulations
  of oblique pulsars},}\ }\href {\doibase 10.1088/2041-8205/801/1/L19}
  {\bibfield  {journal} {\bibinfo  {journal} {Astrophys. J. Lett.}\ }\textbf
  {\bibinfo {volume} {801}},\ \bibinfo {pages} {L19} (\bibinfo {year}
  {2015})}\BibitemShut {NoStop}%
\bibitem [{\citenamefont {{Philippov}}\ \emph {et~al.}(2015)\citenamefont
  {{Philippov}}, \citenamefont {{Cerutti}}, \citenamefont {{Tchekhovskoy}},\
  and\ \citenamefont {{Spitkovsky}}}]{philippov_2015b}%
  \BibitemOpen
  \bibfield  {author} {\bibinfo {author} {\bibfnamefont {A.~A.}\ \bibnamefont
  {{Philippov}}}, \bibinfo {author} {\bibfnamefont {B.}~\bibnamefont
  {{Cerutti}}}, \bibinfo {author} {\bibfnamefont {A.}~\bibnamefont
  {{Tchekhovskoy}}}, \ and\ \bibinfo {author} {\bibfnamefont {A.}~\bibnamefont
  {{Spitkovsky}}},\ }\bibfield  {title} {\enquote {\bibinfo {title}
  {{\textit{Ab initio} pulsar magnetosphere: the role of General
  Relativity}},}\ }\href {\doibase 10.1088/2041-8205/815/2/L19} {\bibfield
  {journal} {\bibinfo  {journal} {Astrophys. J. Lett.}\ }\textbf {\bibinfo
  {volume} {815}},\ \bibinfo {pages} {L19} (\bibinfo {year}
  {2015})}\BibitemShut {NoStop}%
\bibitem [{\citenamefont {Chen}, \citenamefont {Cruz},\ and\ \citenamefont
  {Spitkovsky}(2020)}]{chen_2020}%
  \BibitemOpen
  \bibfield  {author} {\bibinfo {author} {\bibfnamefont {A.~Y.}\ \bibnamefont
  {Chen}}, \bibinfo {author} {\bibfnamefont {F.}~\bibnamefont {Cruz}}, \ and\
  \bibinfo {author} {\bibfnamefont {A.}~\bibnamefont {Spitkovsky}},\ }\bibfield
   {title} {\enquote {\bibinfo {title} {Filling the magnetospheres of weak
  pulsars},}\ }\href {\doibase 10.3847/1538-4357/ab5c20} {\bibfield  {journal}
  {\bibinfo  {journal} {Astrophys. J.}\ }\textbf {\bibinfo {volume} {889}},\
  \bibinfo {pages} {69} (\bibinfo {year} {2020})}\BibitemShut {NoStop}%
\bibitem [{\citenamefont {Gu\'epin}, \citenamefont {Cerutti},\ and\
  \citenamefont {Kotera}(2020)}]{guepin_2020}%
  \BibitemOpen
  \bibfield  {author} {\bibinfo {author} {\bibfnamefont {C.}~\bibnamefont
  {Gu\'epin}}, \bibinfo {author} {\bibfnamefont {B.}~\bibnamefont {Cerutti}}, \
  and\ \bibinfo {author} {\bibfnamefont {K.}~\bibnamefont {Kotera}},\ }\href
  {\doibase 10.1051/0004-6361/201936816} {\bibfield  {journal} {\bibinfo
  {journal} {Astron. Astrophys.}\ }\textbf {\bibinfo {volume} {635}},\ \bibinfo
  {pages} {A138} (\bibinfo {year} {2020})}\BibitemShut {NoStop}%
\bibitem [{\citenamefont {Ben{\'{a}}{\v{c}}ek}\ \emph
  {et~al.}(2021)\citenamefont {Ben{\'{a}}{\v{c}}ek}, \citenamefont
  {Mu{\~{n}}oz}, \citenamefont {Manthei},\ and\ \citenamefont
  {Büchner}}]{benacek_2021a}%
  \BibitemOpen
  \bibfield  {author} {\bibinfo {author} {\bibfnamefont {J.}~\bibnamefont
  {Ben{\'{a}}{\v{c}}ek}}, \bibinfo {author} {\bibfnamefont {P.~A.}\
  \bibnamefont {Mu{\~{n}}oz}}, \bibinfo {author} {\bibfnamefont {A.~C.}\
  \bibnamefont {Manthei}}, \ and\ \bibinfo {author} {\bibfnamefont
  {J.}~\bibnamefont {Büchner}},\ }\bibfield  {title} {\enquote {\bibinfo
  {title} {Radio emission by soliton formation in relativistically hot
  streaming pulsar pair plasmas},}\ }\href {\doibase 10.3847/1538-4357/ac0338}
  {\bibfield  {journal} {\bibinfo  {journal} {Astrophys. J.}\ }\textbf
  {\bibinfo {volume} {915}},\ \bibinfo {pages} {127} (\bibinfo {year}
  {2021})}\BibitemShut {NoStop}%
\bibitem [{\citenamefont {Ben{\'{a}}{\v{c}}ek}, \citenamefont {Mu{\~{n}}oz},\
  and\ \citenamefont {Büchner}(2021)}]{benacek_2021b}%
  \BibitemOpen
  \bibfield  {author} {\bibinfo {author} {\bibfnamefont {J.}~\bibnamefont
  {Ben{\'{a}}{\v{c}}ek}}, \bibinfo {author} {\bibfnamefont {P.~A.}\
  \bibnamefont {Mu{\~{n}}oz}}, \ and\ \bibinfo {author} {\bibfnamefont
  {J.}~\bibnamefont {Büchner}},\ }\bibfield  {title} {\enquote {\bibinfo
  {title} {Bunch expansion as a cause for pulsar radio emissions},}\ }\href
  {\doibase 10.3847/1538-4357/ac2c64} {\bibfield  {journal} {\bibinfo
  {journal} {Astrophys. J.}\ }\textbf {\bibinfo {volume} {923}},\ \bibinfo
  {pages} {99} (\bibinfo {year} {2021})}\BibitemShut {NoStop}%
\bibitem [{\citenamefont {Kelner}, \citenamefont {Prosekin},\ and\
  \citenamefont {Aharonian}(2015)}]{kelner_2015}%
  \BibitemOpen
  \bibfield  {author} {\bibinfo {author} {\bibfnamefont {S.~R.}\ \bibnamefont
  {Kelner}}, \bibinfo {author} {\bibfnamefont {A.~Y.}\ \bibnamefont
  {Prosekin}}, \ and\ \bibinfo {author} {\bibfnamefont {F.~A.}\ \bibnamefont
  {Aharonian}},\ }\href {\doibase 10.1088/0004-6256/149/1/33} {\bibfield
  {journal} {\bibinfo  {journal} {Astrophys. J.}\ }\textbf {\bibinfo {volume}
  {149}},\ \bibinfo {pages} {33} (\bibinfo {year} {2015})}\BibitemShut
  {NoStop}%
\bibitem [{\citenamefont {{Del Gaudio}}(2020)}]{delgaudio_2020}%
  \BibitemOpen
  \bibfield  {author} {\bibinfo {author} {\bibfnamefont {F.}~\bibnamefont {{Del
  Gaudio}}},\ }\href@noop {} {Ph.D. thesis},\ \bibinfo  {school} {Instituto
  Superior T\'{e}cnico} (\bibinfo {year} {2020})\BibitemShut {NoStop}%
\bibitem [{\citenamefont {Arons}\ and\ \citenamefont
  {Scharlemann}(1979)}]{arons_1979}%
  \BibitemOpen
  \bibfield  {author} {\bibinfo {author} {\bibfnamefont {J.}~\bibnamefont
  {Arons}}\ and\ \bibinfo {author} {\bibfnamefont {E.~T.}\ \bibnamefont
  {Scharlemann}},\ }\bibfield  {title} {\enquote {\bibinfo {title} {Pair
  formation above pulsar polar caps: structure of the low altitude acceleration
  zone},}\ }\href {\doibase 10.1086/157250} {\bibfield  {journal} {\bibinfo
  {journal} {Astrophys. J.}\ }\textbf {\bibinfo {volume} {231}},\ \bibinfo
  {pages} {854--879} (\bibinfo {year} {1979})}\BibitemShut {NoStop}%
\bibitem [{\citenamefont {Corless}\ \emph {et~al.}(1996)\citenamefont
  {Corless}, \citenamefont {Gonnet}, \citenamefont {Hare}, \citenamefont
  {Jeffrey},\ and\ \citenamefont {Knuth}}]{corless_1996}%
  \BibitemOpen
  \bibfield  {author} {\bibinfo {author} {\bibfnamefont {R.~M.}\ \bibnamefont
  {Corless}}, \bibinfo {author} {\bibfnamefont {G.~H.}\ \bibnamefont {Gonnet}},
  \bibinfo {author} {\bibfnamefont {D.~E.~G.}\ \bibnamefont {Hare}}, \bibinfo
  {author} {\bibfnamefont {D.~J.}\ \bibnamefont {Jeffrey}}, \ and\ \bibinfo
  {author} {\bibfnamefont {D.~E.}\ \bibnamefont {Knuth}},\ }\href {\doibase
  10.1007/BF02124750} {\bibfield  {journal} {\bibinfo  {journal} {Adv. Comput.
  Math.}\ }\textbf {\bibinfo {volume} {5}},\ \bibinfo {pages} {329–--359}
  (\bibinfo {year} {1996})}\BibitemShut {NoStop}%
\bibitem [{\citenamefont {Fonseca}\ \emph {et~al.}(2002)\citenamefont {Fonseca}
  \emph {et~al.}}]{fonseca_2002}%
  \BibitemOpen
  \bibfield  {author} {\bibinfo {author} {\bibfnamefont {R.~A.}\ \bibnamefont
  {Fonseca}} \emph {et~al.},\ }\bibfield  {title} {\enquote {\bibinfo {title}
  {{OSIRIS}: A three-dimensional, fully relativistic particle in cell code for
  modeling plasma based accelerators},}\ }in\ \href@noop {} {\emph {\bibinfo
  {booktitle} {Computational Science --- ICCS 2002}}},\ \bibinfo {editor}
  {edited by\ \bibinfo {editor} {\bibfnamefont {P.~M.~A.}\ \bibnamefont
  {Sloot}}, \bibinfo {editor} {\bibfnamefont {A.~G.}\ \bibnamefont {Hoekstra}},
  \bibinfo {editor} {\bibfnamefont {C.~J.~K.}\ \bibnamefont {Tan}}, \ and\
  \bibinfo {editor} {\bibfnamefont {J.~J.}\ \bibnamefont {Dongarra}}}\
  (\bibinfo  {publisher} {Springer Berlin Heidelberg},\ \bibinfo {address}
  {Berlin, Heidelberg},\ \bibinfo {year} {2002})\ pp.\ \bibinfo {pages}
  {342--351}\BibitemShut {NoStop}%
\bibitem [{\citenamefont {Fonseca}\ \emph {et~al.}(2008)\citenamefont
  {Fonseca}, \citenamefont {Martins}, \citenamefont {Silva}, \citenamefont
  {Tonge}, \citenamefont {Tsung},\ and\ \citenamefont {Mori}}]{fonseca_2008}%
  \BibitemOpen
  \bibfield  {author} {\bibinfo {author} {\bibfnamefont {R.~A.}\ \bibnamefont
  {Fonseca}}, \bibinfo {author} {\bibfnamefont {S.~F.}\ \bibnamefont
  {Martins}}, \bibinfo {author} {\bibfnamefont {L.~O.}\ \bibnamefont {Silva}},
  \bibinfo {author} {\bibfnamefont {J.~W.}\ \bibnamefont {Tonge}}, \bibinfo
  {author} {\bibfnamefont {F.~S.}\ \bibnamefont {Tsung}}, \ and\ \bibinfo
  {author} {\bibfnamefont {W.~B.}\ \bibnamefont {Mori}},\ }\href {\doibase
  10.1088/0741-3335/50/12/124034} {\bibfield  {journal} {\bibinfo  {journal}
  {Plasma Phys. Control. Fusion}\ }\textbf {\bibinfo {volume} {50}},\ \bibinfo
  {pages} {124034} (\bibinfo {year} {2008})}\BibitemShut {NoStop}%
\bibitem [{\citenamefont {Cruz}, \citenamefont {Grismayer},\ and\ \citenamefont
  {Silva}(2021{\natexlab{b}})}]{cruz_2021c}%
  \BibitemOpen
  \bibfield  {author} {\bibinfo {author} {\bibfnamefont {F.}~\bibnamefont
  {Cruz}}, \bibinfo {author} {\bibfnamefont {T.}~\bibnamefont {Grismayer}}, \
  and\ \bibinfo {author} {\bibfnamefont {L.~O.}\ \bibnamefont {Silva}},\
  }\bibfield  {title} {\enquote {\bibinfo {title} {Kinetic instability in
  inductively oscillatory plasma equilibrium},}\ }\href {\doibase
  10.1103/PhysRevE.103.L051201} {\bibfield  {journal} {\bibinfo  {journal}
  {Phys. Rev. E}\ }\textbf {\bibinfo {volume} {103}},\ \bibinfo {pages}
  {L051201} (\bibinfo {year} {2021}{\natexlab{b}})}\BibitemShut {NoStop}%
\bibitem [{\citenamefont {{Goldreich}}\ and\ \citenamefont
  {{Julian}}(1969)}]{goldreich_julian_1969}%
  \BibitemOpen
  \bibfield  {author} {\bibinfo {author} {\bibfnamefont {P.}~\bibnamefont
  {{Goldreich}}}\ and\ \bibinfo {author} {\bibfnamefont {W.~H.}\ \bibnamefont
  {{Julian}}},\ }\bibfield  {title} {\enquote {\bibinfo {title} {Pulsar
  electrodynamics},}\ }\href {\doibase 10.1086/150119} {\bibfield  {journal}
  {\bibinfo  {journal} {Astrophys. J.}\ }\textbf {\bibinfo {volume} {157}},\
  \bibinfo {pages} {869} (\bibinfo {year} {1969})}\BibitemShut {NoStop}%
\bibitem [{\citenamefont {Beloborodov}(2008)}]{beloborodov_2008}%
  \BibitemOpen
  \bibfield  {author} {\bibinfo {author} {\bibfnamefont {A.~M.}\ \bibnamefont
  {Beloborodov}},\ }\bibfield  {title} {\enquote {\bibinfo {title} {Polar-cap
  accelerator and radio emission from pulsars},}\ }\href {\doibase
  10.1086/590079} {\bibfield  {journal} {\bibinfo  {journal} {Astrophys. J.
  Lett.}\ }\textbf {\bibinfo {volume} {683}},\ \bibinfo {pages} {L41--L44}
  (\bibinfo {year} {2008})}\BibitemShut {NoStop}%
\end{thebibliography}%



\end{document}